\global\def\draftcontrol{0}

   \def\versionno{ stability }

\catcode`\@=11

\expandafter\ifx\csname draftcontrol\endcsname\relax\global\def\draftcontrol{0} 
\fi 

{\count255=\time\divide\count255 by 60 
\xdef\hourmin{\number\count255} 
\multiply\count255 by-60\advance\count255 by\time 
\xdef\hourmin{\hourmin:\ifnum\count255<10 0\fi\the\count255}} 
\def\draftdate{\number\month/\number\day/\number\year\ \ \ \hourmin } 

\newcommand\makepapertitle{\par

  \begingroup 
    \renewcommand\thefootnote{\@fnsymbol\c@footnote}%
    \def\@makefnmark{\rlap{\@textsuperscript{\normalfont\@thefnmark}}}%
    \long\def\@makefntext##1{\parindent 1em\noindent 
            \hb@xt@1.8em{%
                \hss\@textsuperscript{\normalfont\@thefnmark}}##1}%
     \newpage 
     \global\@topnum\z@   
     \@makepapertitle 
     \thispagestyle{empty}\@thanks 
  \endgroup 
  \setcounter{footnote}{0}%
  \global\let\thanks\relax 
  \global\let\makepapertitle\relax 
  \global\let\@makepapertitle\relax 
  \global\let\@thanks\@empty 
  \global\let\@author\@empty 
  \global\let\@date\@empty 
  \global\let\@title\@empty 
  \global\let\title\relax 
  \global\let\author\relax 
  \global\let\date\relax 
  \global\let\and\relax 
  \def\version{\let\version\@version\@gobble} 
} 
\def\@makepapertitle{%
  \newpage 
   \ifnum\draftcontrol=1 {} 
   \version\versionno 
   \vskip 5em%
   \else 
   \hfill\hbox to 3cm {\parbox{4cm}{\@pubnum}\hss}%
   \vskip 5em%
   \fi 
   \begin{center}%
   \let \footnote \thanks 
      {\hskip -0\textwidth \hbox to 1\textwidth%
        {\centerline{\Large\bf{\noindent\@title}}}}%
     \vskip 2em%
     {\normalsize
       \lineskip .5em%
       \begin{tabular}[t]{c}%
         \@author 
       \end{tabular}\par}%
     \vskip 1em%
     {\@bstract}%
     \end{center}%
     \vfill
     \@date%
     \vskip 1.5em%
   \par 
} 

\gdef\@pubnum{} 
\def\pubnum#1{%
  \gdef\@pubnum{#1}} 

\gdef\@bstract{} 
\def\Abstract#1{%
  \gdef\@bstract{%
   \parbox{\textwidth-0pc}{%
   \centerline{\bf Abstract}\penalty1000 
   \noindent
   \renewcommand\baselinestretch{1.0} 
   {#1}}} 
} 

\gdef\@email{}
\def\email#1{%
   \gdef\@email{%
   Email: {\tt #1}}
}

\def\ps@paper{\let\@mkboth\@gobbletwo%
     \ifnum\draftcontrol=1 
        \def\@oddfoot{\hbox to \textwidth{\tiny \versionno \hfil\tiny\draftdate}%
        \hskip -\textwidth \hbox to \textwidth{\hfil\rm\thepage\hfil}}%
     \else\def\@oddfoot{\hbox to \textwidth{\hfil\rm\thepage\hfil}} 
     \fi 
     \let\@evenfoot\@oddfoot 
} 

\def\body{\clearpage 
          \pagestyle{paper} 
        } 
\newenvironment{acknowledgments}{%
\vskip 3.25ex 
\noindent {\bf Acknowledgments} 
} 

\def\@version#1{\ifnum\draftcontrol=1 
\typeout{}\typeout{#1}\typeout{} 
\vskip3mm\centerline{\hbox{\fbox{\normalsize{\tt DRAFT -- #1 -- } 
                   {\draftdate}}}}\vskip3mm 
\fi} 
\let\version\@version 
\long\def\eqlabel#1{\ifnum\draftcontrol=1 
                    \tag@false  
                    \tag*{(\theequation) \hbox to -0.2cm{\hspace{0cm}\small{#1}\hss}} 
                    \refstepcounter{equation}  
                    \edef\@currentlabel{\theequation} 
                    \ltx@label{#1}          
                    \else 
                    \label{#1} 
                    \fi 
                    } 
\let\st@bibitem\@bibitem 
\let\st@lbibitem\@lbibitem 
\ifnum\draftcontrol=1 
  \def\@bibitem#1{%
    \st@bibitem{#1}\a@@label{#1}\ignorespaces} 
  \def\@lbibitem[#1]#2{%
    \st@lbibitem[#1]{#2}\a@@label{#2}\ignorespaces} 
  \def\a@@label#1{%
    \gdef\a@lab{\smash{\normalfont\small#1}} 
    \ifvmode 
      \if@inlabel 
        \global\setbox\@labels\hbox{%
          \llap{\a@lab\let\a@lab\relax 
                \kern\@totalleftmargin\kern\marginparsep}%
          \box\@labels}%
      \fi 
    \fi} 
\fi 

\documentclass[12pt,letterpaper]{article} 

\usepackage[ps,dvips,matrix,arrow,frame,import,curve,color]{xy}
\usepackage{amsmath,bm,amsfonts,amssymb,array,calc,amsthm,rotating}
\usepackage[nosort]{cite} 
\usepackage{graphicx}

\renewcommand\baselinestretch{1.25} 
\renewcommand\section{\@startsection {section}{1}{\z@}%
                                   {-3.5ex \@plus -1ex \@minus -.2ex}%
                                   {2.3ex \@plus.2ex}%
                                   {\normalfont\large\bfseries}} 
\renewcommand\subsection{\@startsection{subsection}{2}{\z@}%
                                   {-3.25ex\@plus -1ex \@minus -.2ex}%
                                   {1.5ex \@plus .2ex}%
                                   {\normalfont\normalsize\bfseries}} 
\renewcommand\subsubsection{\@startsection{subsubsection}{3}{\z@}%
                                   {-3.25ex\@plus -1ex \@minus -.2ex}%
                                   {1.5ex \@plus .2ex}%
                                   {\normalfont\normalsize\it}} 
\renewcommand\paragraph{\@startsection{paragraph}{4}{\z@}%
                                   {-3.25ex\@plus -1ex \@minus -.2ex}%
                                   {1.5ex \@plus .2ex}%
                                   {\normalfont\normalsize\bf}} 
\renewcommand\subparagraph{\@startsection{subparagraph}{5}{\z@}%
                                   {-1.25ex\@plus -1ex \@minus -.2ex}%
                                   {0ex \@plus .2ex}%
                                   {\normalfont\normalsize\it}} 

\numberwithin{equation}{section}

\long\def\@makecaption#1#2{%
  \vskip\abovecaptionskip
  \sbox\@tempboxa{{\bf #1:} #2}%
  \ifdim \wd\@tempboxa >\hsize
    {\small\bf #1:} {\small #2}\par
  \else
    \global \@minipagefalse
    \hb@xt@\hsize{\hfil\box\@tempboxa\hfil}%
  \fi
  \vskip\belowcaptionskip}

\renewcommand*\l@section[2]{%
  \ifnum \c@tocdepth >\z@
    \addpenalty\@secpenalty
    \addvspace{.5em \@plus\p@}%
    \setlength\@tempdima{1.5em}%
    \begingroup
      \parindent \z@ \rightskip \@pnumwidth
      \parfillskip -\@pnumwidth
      \leavevmode \bfseries
      \advance\leftskip\@tempdima
      \hskip -\leftskip
      #1\nobreak\hfil \nobreak\hb@xt@\@pnumwidth{\hss #2}\par
    \endgroup
  \fi}
\renewcommand*\l@subsection{\addvspace{.0em \@plus\p@}\@dottedtocline{2}{1.5em}{2.3em}}
\renewcommand*\l@subsubsection{\addvspace{-.2em \@plus\p@}\@dottedtocline{3}{3.8em}{3.2em}}

\def\ie{{\it i.e.}} 
\def\eg{{\it e.g.}} 
\def\etc{{\it etc.}}

\def\revise#1       {\raisebox{-0em}{\rule{3pt}{1em}}%
                     \marginpar{\raisebox{.5em}{\vrule width3pt\ 
                     \vrule width0pt height 0pt depth0.5em 
                     \hbox to 0cm{\hspace{0cm}{%
                     \parbox[t]{4em}{\raggedright\footnotesize{#1}}}\hss}}}}

\def\calb         {{\cal B}} 
\def\calc         {{\cal C}}

\def\calg         {{\cal G}} 
 
\def\calh         {{\cal H}} 
 
\def\calj         {{\cal J}}

\def\calm         {{\cal M}} 
\def\caln         {{\cal N}} 
\def\calo         {{\cal O}}

\def\calr         {{\cal R}} 
\def\cals         {{\cal S}}

\def\calv         {{\cal V}}

\def\complex      {{\mathbb C}} 
 
\def\projective   {{\mathbb P}} 
\def\rationals    {{\mathbb Q}} 
\def\reals        {{\mathbb R}} 
\def\zet          {{\mathbb Z}} 

\def\del          {\partial} 
 
\def\ee           {{\it e}} 
\def\ii           {{\it i}} 
 
\def\tr           {{\rm Tr}} 
\def\Re           {{\rm Re\hskip0.1em}} 
\def\Im           {{\rm Im\hskip0.1em}} 
\def\id           {{\rm id}}

\def\sqr#1#2{{\vcenter{\vbox{\hrule height.#2pt   
 \hbox{\vrule width.#2pt height#1pt \kern#1pt 
 \vrule width.#2pt}\hrule height.#2pt}}}}

\def\Hom{{\rm Hom}}
\def\U{{\it U}}
\def\GL{{\it GL}}
\def\Ker{{\rm Ker}}
\def\Im{{\rm Im}}
\def\Coker{{\rm Coker}}

\def\HH{\mathfrak{H}}
\def\CC{\mathfrak{C}}

\def\gg{\mathfrak{g}}

\def\str{{\mathop{\rm Str}}}

\def\DD{{\bf D}}          
\def\MCM{{\rm MCM}}       
\def\free{{\rm DG}}       
\def\mf{{\rm MF}}          
\def\MF{\mathfrak{MF}}    

\def\ch{{\rm ch}}
\def\td{{\rm Td}}

\def\res{{\rm Res}}

\def\calrloc{\calr_{\mathfrak m}}
\def\tcalrloc{{\tilde\calr}_{\mathfrak m}}

\catcode`\@=12 

\begin{document} 


\title{Stability of Landau-Ginzburg branes}

\pubnum{%
hep-th/0412274}
\date{December 2004}

\author{
Johannes Walcher \\[0.4cm]
\it School of Natural Sciences, Institute for Advanced Study\\
\it Princeton, New Jersey, USA
}

\Abstract{
We evaluate the ideas of $\Pi$-stability at the Landau-Ginzburg
point in moduli space of compact Calabi-Yau manifolds, using 
matrix factorizations to B-model the topological D-brane category. 
The standard requirement of unitarity at the IR fixed point is argued
to lead to a notion of ``R-stability'' for matrix factorizations of 
quasi-homogeneous LG potentials. The D$0$-brane on the quintic at the
Landau-Ginzburg point is not obviously unstable. Aiming to relate 
R-stability to a moduli space problem, we then study the action of 
the gauge group of similarity transformations on matrix factorizations. 
We define a naive moment map-like flow on the gauge orbits and use it 
to study boundary flows in several examples. Gauge transformations of 
non-zero degree play an interesting role for brane-antibrane annihilation. 
We also give a careful exposition of the grading of the Landau-Ginzburg 
category of B-branes, and prove an index theorem for matrix 
factorizations.
}

\makepapertitle

\body

\version\versionno

\vskip 1em

\tableofcontents
\newpage

\section{Introduction}

The main purpose of this work is to develop a stability condition,
to be called ``R-stability'', on the triangulated category of 
matrix factorizations describing D-branes at the Landau-Ginzburg 
point $p_{\rm LG}$ in the K\"ahler moduli space $\calm_k$ of a 
compact Calabi-Yau manifold $X$. The proposal is motivated by 
physical considerations similar to the ones leading to the notion 
of $\Pi$-stability on the derived category of coherent sheaves 
$\DD(X)$, which describes the variation of the spectrum of B-type
BPS branes over $\calm_k$. In fact, the notion of R-stability can 
be thought of as the specialization of $\Pi$-stability to $p_{\rm LG}$. 
It is expected, however, that R-stability should be intrinsic to the
Landau-Ginzburg model and does in principle not depend on 
knowledge of the stable spectrum elsewhere in $\calm_k$.

In this paper, section \ref{review} is a brief review of the relevant 
aspects of $\Pi$-stability that we want to abstract to the 
Landau-Ginzburg model. Section \ref{factor} contains the basic 
definitions related to matrix factorizations. Section \ref{graded} 
explains how quasi-homogeneous matrix factorizations can be, 
first $\rationals$-, then $\zet$-graded. Section \ref{index} is 
a somewhat independent unit concerned with the RR charges of 
matrix factorizations in string theory and an index theorem. 
Section \ref{stabcon} gives a preliminary definition of 
R-stability and partial answers to the difficulties in relating
it to the action of the gauge group on matrix factorizations. This 
general discussion is then applied in section \ref{examples}
and the proposal shown to work well in several relevant examples.
Section \ref{summary} gives a summary.

\section{Review of $\Pi$-stability}
\label{review}

$\Pi$-stability was introduced in \cite{dfr,douglas}, and 
further sharpened and tested in \cite{asla,asdo}. It was
subsequently abstracted into a precise mathematical definition
of stability condition on triangulated categories in 
\cite{bridgeland}. We refer to these works for the categorical
aspects of $\Pi$-stability, as well as to Aspinwall's review 
\cite{aspreview} for more extensive background material. 
Instead, we begin with a slightly personal review of the 
worldsheet origin of $\Pi$-stability, following Douglas 
\cite{douglas}.

The basic physical intuition is quite simple. Consider a fixed 
2-dimensional conformally invariant string worldsheet quantum field
theory $\calc$ defining a closed string background. By definition, 
a D-brane in this background is a conformally invariant boundary 
condition, $\calb$, for $\calc$. A popular way to define $\calc$'s 
and $\calb$'s is as IR fixed points of bulk or boundary RG flows, 
induced by turning on a relevant operator $\calo$ in a known 
bulk or boundary theory. Such a UV description is ``stable'' if it 
flows to a theory in the infrared which is ``acceptable'' in the 
sense of, \eg, having the right central charge, being unitary, 
\etc. Finding necessary and/or sufficient stability conditions 
on $\calo$ is in general a very hard question.

A situation in which more can be said is when one requires bulk 
and boundary theories to preserve $\caln=2$ supersymmetry with a
non-anomalous $\U(1)$ R-symmetry, so that the chiral algebra 
underlying $\calc$ and $\calb$ will contain the $\caln=2$ 
superconformal algebra. A necessary condition on acceptable 
$\calb$'s is that the R-charges of all open string NS chiral 
primary operators satisfy the unitarity constraint \cite{chiral}
\begin{equation}
0\le q\le \hat c \,,
\eqlabel{unitarity}
\end{equation}
where $\hat c$ is the central charge of the superconformal algebra.
Often, $\hat c$ and $q$ can be determined in the UV and the equation 
\eqref{unitarity} therefore provides a stability condition in the above 
sense.

The ideas of $\Pi$-stability in fact go further than \eqref{unitarity}. 
Assume that $\hat c$ and the R-charges of $\calc$ are all integral. The
chiral algebra of $\calc$ then contains, in addition to the $\caln=2$
superconformal algebra, the (square of the) spectral flow operator $\cals$.
One can then contemplate imposing a boundary condition of the form
\begin{equation}
\cals_L = \ee^{\ii\pi\varphi} \cals_R \,,
\eqlabel{sglue}
\end{equation}
involving an arbitrary phase, $\varphi$. Standard conformal field theory 
arguments\footnote{Using the doubling trick, one transports $\cals^2$ 
around an open string vertex operator inserted at the boundary of the 
worldsheet, and notes that the total monodromy of $\cals^2$, which 
evaluates to the difference of phases, measures the $U(1)$ charge of 
the operator up to an integer.} then show that the R-charges of an 
open string spanning between two branes $\calb$ and $\calb'$ (with 
phase $\varphi$ and $\varphi'$) satisfy
\begin{equation}
q = \varphi' -\varphi \bmod \zet \,.
\eqlabel{qmod}
\end{equation}
If we bosonize the left- and right-moving $\U(1)$ currents in terms
of two canonically normalized chiral bosons $\phi_L$ and $\phi_R$,
the spectral flow operators are $\cals_{L,R}=\ee^{\ii\frac
{\sqrt{\hat c}}2 \phi_{L,R}}$, and we can visualize the boundary 
condition \eqref{sglue} as Dirichlet or Neumann boundary condition on 
the compact boson $\phi=\phi_L\pm\phi_R$ with radius $\sqrt{\hat c}$. 
(The sign depending on which side (A or B) of the mirror one chooses 
to present the conformal field theory.) Of course, the equation
$\ee^{\ii\frac{\sqrt{\hat c}}2 \phi_L} = \ee^{\ii\pi\varphi}
\ee^{\ii\frac{\sqrt{\hat c}}2 \phi_R}$ leaves a $\hat c$-fold
ambiguity on the position/Wilson line of the boundary condition
on $\phi$. In such a picture \cite{douglas}, the strings stretching 
between the different images correspond to the different values of $q$ 
in \eqref{qmod}.

We emphasize that, in conformal field theory, $\varphi$ is defined
as a real number modulo even integers. We should also like to stress 
that $\varphi$ is, in general, independent of the phases appearing in 
the boundary condition on the $\caln=2$ currents, as in, $G_L^{\pm} = 
\ee^{\pm \ii\alpha}G_R^{\pm}$, $G_L^{\pm} = \ee^{\pm\ii\beta}G_R^{\mp}$ 
for A- and B-type, respectively. (See, \eg, \cite{fsw} for a BCFT
discussion of this.) $\varphi$ determines which $\caln=1$ spacetime 
supersymmetry is preserved by the brane, and can be different for 
different branes. On the other hand, the phases appearing in the 
boundary condition on the $\caln=2$ currents determine which 
$\caln=1$ worldsheet supersymmetry is preserved. This is a gauge 
symmetry and has to be the same for all branes.

Now recall that an $\caln=2$ field theory (conformal or not) with a 
conserved $\U(1)$ R-current can be twisted to a topological theory.
As anticipated in \cite{icm}, and by now well appreciated in the 
physics literature, the set of branes in the topological theory 
together with open strings between them carries the algebraic structure 
of a ``triangulated category'' (plus more). Two important pieces
of structure are, firstly, the so-called ``distinguished triangles'', 
such as
\begin{equation}
\xymatrix@R=1cm@C=0.7cm{
& B \ar[dr]^{S_1} & \\
\ar[ur]^{S_2} B_2 && \ar@{-->}[ll]^{T} B_1 
}
\eqlabel{triangle}
\end{equation}
which expresses the fact that the topological brane $B$ can be
obtained as a topological bound state of the two branes $B_1$ and
$B_2$ by condensing the ``topological tachyon'' $T$ on the base of
the triangle. Secondly, a triangulated category has a so-called
shift functor, which in physics terms sends a brane $B$ to a
copy of its antibrane $B[1]$.

In relating the physical to the topological theory, one chooses a 
lift of the phase $\varphi$ to a real number called ``grade'' and 
identifies the ghost number $n$ of open strings as the integer 
appearing in \eqref{qmod}, \ie,
\begin{equation}
n = q + \varphi  - \varphi' \,.
\eqlabel{qunmod}
\end{equation}
Consequently, for every physical brane $\calb$ there are in fact
an infinite number of topological branes $B[m]$ whose grade differs
by an integers. Shifting the grade shifts the ghost number by 
integers, and hence modifies the topological theory. On the other 
hand, the topological theory is unaware of the unitarity constraint 
\eqref{unitarity}. In particular, the topological theory is 
independent of changes of $q$ and $\varphi$. This decoupling of 
the topological theory from the variation of $q$ with (part of) the 
moduli is one of the central ideas underlying $\Pi$-stability.

$\Pi$-stability, then, is designed to decide when the bound
state formation described in the topological theory by
triangles such as \eqref{triangle} is stable in the physical
theory, and thereby provides a picture of the spectrum of BPS
branes in some given closed string background. Let us, for 
concreteness, focus on the case of B-type D-branes on a Calabi-Yau
manifold $X$. In that case, the topological branes are objects of 
the derived category of coherent sheaves, $\DD(X)$, of the algebraic
variety underlying $X$. $\DD(X)$ depends only on the complex
structure of $X$, and is independent of the K\"ahler moduli. Within 
this category of topological branes, the set of stable branes, 
conjectured to flow to BPS branes in the physical theory, varies 
over the stringy K\"ahler moduli space $\calm_k$ of $X$.
Essentially, one follows the continuous variation of the phases
$\varphi$, and hence of $\U(1)$ R-charges of open strings, over 
$\calm_k$. Charges leaving/entering the unitarity bound 
\eqref{unitarity} signal loss/gain of stable branes, with decay
and bound state formation described by the triangles
\eqref{triangle}.

For the details of this construction, consistency with monodromies
in $\calm_k$, and a lot of examples, see ref.\ \cite{aspreview}.
One peculiar aspect of the story is that $\Pi$-stability really 
only describes the {\it changes} of the BPS spectrum as
one moves around in $\calm_k$. To determine the spectrum at any 
given point $p$ of $\calm(X)$, one has to know the spectrum at some 
distinguished point $p_0$ and then follow it to $p$ using 
$\Pi$-stability. 

One natural choice for basepoint is large volume, $p_{\rm LV}$,
in the compactification of $\calm_k$. At $p_{\rm LV}$,
$\Pi$-stability reduces to $\mu$-stability for the Abelian 
category of coherent sheaves on $X$ \cite{dfr}. Although $\mu$-stability
does not extend over an open neighborhood of $p_{\rm LV}$ (and
hence does not allow determining the complete BPS spectrum there), 
it is at present the only useful handle on the spectrum elsewhere 
in $\calm_k$.

Another special point, which one expects exists when $X$ is a 
{\it non-compact} Calabi-Yau manifold, is the so-called ``orbifold 
point'' $p_{\rm O}$ in $\calm$. Even if $X$ is not the resolution of an 
actual orbifold singularity, on may define $p_{\rm O}$ as a point in 
$\calm_k$ at which the phases of all branes are aligned. In such a 
situation, determining the BPS spectrum is a problem of solving F- and 
D-flatness conditions in a supersymmetric quiver gauge theory, as argued 
in \cite{dfr}. More rigorously, Aspinwall shows in \cite{aspinwall}
that in an open neighborhood of such an orbifold point, $\Pi$-stability 
reduces to $\theta$-stability for the Abelian category of quiver 
representations in the sense of King \cite{king}.

Such a point at which all phases align is expected not to exist in 
the moduli space of a generic compact Calabi-Yau model. The closest 
one can get seems to be the Landau-Ginzburg point, which resembles 
ordinary orbifolds in the appearance of a discrete quantum symmetry, 
but with the important difference that not all phases of branes 
are aligned. The purpose of the present paper, pursuing a suggestion
made in \cite{howa,strings}, is to investigate the ideas underlying 
$\Pi$-stability at the Landau-Ginzburg orbifold point in the 
K\"ahler moduli space of compact Calabi-Yau manifolds, using the 
recently introduced description of the topological category using 
matrix factorizations. We now turn to explaining various (old and 
new) aspects of matrix factorizations, and pick up the stability 
discussion in section \ref{stabcon}.

\section{Matrix factorizations}
\label{factor}

Let $W\in \calr=\complex[x_1,\ldots,x_r]$ be a polynomial.
To keep things simple, we will assume throughout that $W$ has
an isolated critical point at the origin $x_i=0$.
A {\it matrix factorization} (of dimension $N$) of $W$ is a 
pair of square matrices $f,g\in {\rm Mat}(N\times N,\calr)$ 
with polynomial entries satisfying 
\begin{equation} 
f g = g f = W \cdot {\rm id}_{N\times N}  \,.
\eqlabel{mafa}
\end{equation} 

A matrix factorization is called {\it reduced} if all entries
of $f$ and $g$ have no constant term, \ie, $f(0)=g(0)=0$.

Matrix factorizations $(f,g)$ and $(f',g')$ are called 
{\it equivalent} if they are related by a similarity 
transformation
\begin{equation}
U_1 f = f' U_2 \qquad U_2 g = g' U_1
\eqlabel{similarity}
\end{equation}
where $U_1, U_2\in\GL(N,\calr)$ are invertible matrices with 
polynomial entries.

\subsection{Maximal Cohen-Macaulay modules}
\label{MCM}

Matrix factorizations originated in Eisenbud's work \cite{eisen} 
in the context of so-called maximal Cohen-Macaulay modules over local 
rings of hypersurface singularities. See \cite{yoshino,brhe} for some
background. An example of such a ring is given by $\tcalrloc=
\calrloc/(W)$, where $\calrloc=\complex[[x_1,\cdots,x_r]]$ is the 
complete local ring of power series, with maximal ideal 
$\mathfrak m=(x_1,\ldots,x_r)$, and $W$ is a polynomial, 
as above. If $(f,g)$ is a matrix factorization of $W$, consider
the $\calrloc$-module $M=\Coker f$ with the $\calrloc$-free 
resolution
\begin{equation}
0\longrightarrow G \overset{f}\longrightarrow
F \longrightarrow M \longrightarrow 0 \,,
\eqlabel{resolution}
\end{equation}
where $F\cong G\cong (\calrloc)^N$ are rank $N$ free modules.
Since multiplication by $W$ on \eqref{resolution} is homotopic
to zero, $M$ descends to a $\tcalrloc$-module, with the infinite
free resolution
\begin{equation}
\cdots \longrightarrow
\tilde G
\overset{f}{\longrightarrow}
\tilde F
\overset{g}\longrightarrow
\tilde G
\overset{f}\longrightarrow
\tilde F
\longrightarrow M \longrightarrow 0 \,.
\eqlabel{infres}
\end{equation}
with $\tilde F\cong\tilde G\cong (\tcalrloc)^N$.

The resolution \eqref{resolution} being of length one, which 
is the codimension of a hypersurface, makes $M$ into a so-called 
maximal Cohen-Macaulay module (MCM) over $\tcalrloc$ (see 
\cite{yoshino,brhe} for the definitions). Eisenbud's theorem 
\cite{eisen} essentially says that all MCMs over hypersurface rings
come from matrix factorizations. 

The category of Cohen-Macaulay modules \cite{yoshino} will be denoted
by $\MCM(W)$. Objects of $\MCM(W)$ are matrix factorizations 
of $W$ and morphism are morphisms of Cohen-Macaulay modules. In other 
words, a morphism from $(f,g)$ to $(f',g')$ in $\MCM(W)$ is a pair 
of $N'\times N$-dimensional matrices $a$, $b$, with polynomial entries, 
satisfying
\begin{equation}
b g = g' a\qquad a f = f' b \,,
\end{equation}
so that the diagram
\begin{equation}
\xymatrix@C=1cm@R=1cm{
\ar[d]_a \ar[r]^g F & \ar[d]_b \ar[r]^f G & \ar[d]_a F \\
\ar[r]^{g'} F' & \ar[r]^{f'} G'  & F'
}
\eqlabel{hommcm}
\end{equation}
commutes. We will make no direct use of the category $\MCM(W)$,
but have included its definition here since it might play a
role in a precise formulation of R-stability.

\subsection{Triangulated category}

A different category's construction based on matrix factorization
was observed by Kontsevich \cite{kont}. The construction starts
from triples $(M,\sigma,Q)$, where $M$ is a free $\calr=\complex[
x_1,\ldots,x_r]$-module with a $\zet_2$-grading $\sigma$, and $Q$ 
is an odd ($\sigma Q + Q\sigma=0$) endomorphism of $M$ satisfying
\begin{equation}
Q^2 = W \cdot {\rm id}_M \,.
\eqlabel{matfac}
\end{equation}
Decomposing $M=M_0\oplus M_1$ into homogeneous components, with
equal rank $N$, $Q$ can be represented as the matrix
\begin{equation}
Q = \begin{pmatrix} 0 & f \\ g & 0 \end{pmatrix}\,,
\end{equation}
making the relation of \eqref{matfac} to \eqref{mafa} obvious.
The grading is then given by the matrix
\begin{equation}
\sigma = \begin{pmatrix} \id_{N\times N} & 0 \\ 
0 & -\id_{N\times N}\end{pmatrix} \,.
\end{equation}
Let us denote by $\free(W)$ the category which has such triples as 
objects and as morphisms the (even) morphisms of free modules
(forgetting the $Q$'s). The gauge transformations in $\free(W)$ are 
the even automorphisms of $M$ as an $\calr$-module, $\GL^+(2N,\calr)$,
acting as
\begin{equation}
\GL^+(2N,\calr)\ni U =\begin{pmatrix} U_1 & 0\\0&U_2
\end{pmatrix} : Q\mapsto U Q U^{-1}
\end{equation}
with $U_1,U_2\in\GL(N,\calr)$, as in \eqref{similarity}.

The point of the construction \cite{kali1,orlov} is that the 
category $\free(W)$ has the structure of a differential graded
category. This means that morphism spaces $\Hom_\calr(M,M')$
are equipped with an odd differential $D$ acting as a supercommutator
\begin{equation}
\begin{split}
D\Phi &= Q' \Phi - \sigma' \Phi \sigma \; \Phi Q \\
&= \begin{pmatrix} 0 & f'\\g'&0
\end{pmatrix}
\begin{pmatrix} A & B \\ C & D \end{pmatrix}
-
\begin{pmatrix} A & -B \\ -C & D \end{pmatrix} 
\begin{pmatrix} 0 & f \\ g & 0 
\end{pmatrix}
\end{split}
\eqlabel{superco}
\end{equation}
on morphisms
\begin{equation}
\Phi = \begin{pmatrix} A & B \\ C & D \end{pmatrix} 
\end{equation}
in $\free(W)$. One easily checks that $D^2=0$ by the super-Jacobi 
identity. By a general construction \cite{boka,orlov}, one can then 
associate a triangulated category, $\mf(W)$ to $\free(W)$, which 
has the same objects as $\free(W)$ (\ie, triples $(M,\sigma,Q)$),
but in which morphisms are given by the $\zet_2$-graded cohomology 
of $Q$. Thus $\Hom_{\mf(W)}(M,M')= H^*(D) = \Ker D/\Im D$. 
We shall usually embezzle $M$ and $\sigma$, and simply write 
$\Hom_{\mf(W)}(Q,Q')=H^0(Q,Q')\oplus H^1(Q,Q')$ for the morphisms in 
$\mf(W)$. We also write $H^0(Q)$, $H^1(Q)$ for the morphisms from 
$Q$ to itself.

For future reference, let us spell out a few triangulated 
constructions in the language of matrix factorizations.
Firstly, the shift functor $[1]$ is nothing but the reversal
of the $\zet_2$-grading $\sigma\to-\sigma$, or, equivalently, 
the exchange of $f$ and $g$, \ie,
\begin{equation}
Q[1] = \begin{pmatrix} 0 & f \\ g & 0 \end{pmatrix}[1]
= \begin{pmatrix} 0 & g \\ f & 0 \end{pmatrix} \,,
\end{equation}
with $M$ and $\sigma$ fixed. This operation obviously exchanges $H^0$ 
with $H^1$. Secondly, given two matrix factorizations $Q_1$ and $Q_2$ 
and an odd morphism $T\in H^1(Q_1,Q_2)$, we obtain a third factorization 
simply as
\begin{equation}
Q = \begin{pmatrix} Q_1 & 0 \\ T & Q_2 \end{pmatrix}
\eqlabel{cone}
\end{equation}
fitting into the triangle
\begin{equation}
\xymatrix@R=1cm@C=0.7cm{
& Q \ar[dr]^{S_1} & \\
\ar[ur]^{S_2} Q_2 && \ar@{-->}[ll]^{T} Q_1 
} \,,
\eqlabel{tria}
\end{equation}
where
\begin{equation}
S_1 = \begin{pmatrix} 1 & 0 \end{pmatrix}
\qquad
S_2 = \begin{pmatrix} 0  \\  1\end{pmatrix} \,.
\end{equation}
The construction \eqref{cone} is referred to as the ``cone'' over 
the map $T[1]\in H^0(Q_1,Q_2[1])$. 

Let us also note explicitly that the construction of $\mf(W)$
implies in particular that we identify matrix factorizations
which differ by the direct addition of the trivial factorization
$f=1$, $g=W$,
\begin{equation}
Q \equiv Q \oplus 
\begin{pmatrix} 0 & 1 \\ W & 0 \end{pmatrix}
\equiv Q \oplus
\begin{pmatrix} 0 & W \\ 1 & 0 \end{pmatrix}
\eqlabel{trivial}
\end{equation}
This identification occurs because adding the trivial factorization
does not affect the cohomology of $D$ between $Q$ and any other
factorization $Q'$.

\subsection{Relation to $\caln=2$ Landau-Ginzburg model}

Orbifoldized $\caln=2$ Landau-Ginzburg models \cite{lgorb} are known 
\cite{gvw,glsm} to describe the small-volume continuation of 
Calabi-Yau sigma models, see \cite{horietal} for the background.
(LG also describe, in particular, the mirrors of CY sigma models, 
as well as the mirrors of toric Fano and non-compact Calabi-Yau
manifolds, but this will not be important here. We will stay on 
the B-side throughout.)

In the bulk, LG models are characterized by the worldsheet 
superpotential $W$, such as the polynomial we have been studying 
in this section. The $x_i$ are $\caln=2$ chiral field variables,
whose interaction is described by $W$. The kinetic term for the
$x_i$ is described by a K\"ahler potential $K(x_i,\bar x_i)$,
and is usually ignored in the discussion of LG models because
it does not affect topological quantities such as the chiral
ring. What is more, in the quasi-homogeneous case, it is actually
conjectured that there is a K\"ahler potential, uniquely 
determined by the superpotential, such that the associated model 
is conformal. This K\"ahler potential can be reached by RG flow
along which $W$ is unchanged by non-renormalization theorems.

When adding boundaries to the worldsheet of an $\caln=2$ LG model, 
the supersymmetry variation of the superpotential exhibits a peculiar
boundary term, whose non-vanishing is known as the Warner problem 
\cite{warner,gjs,hiv,linear,hklm}. Following a proposal of Kontsevich, 
it was shown in \cite{kali1,bhls,hori} that matrix factorizations of 
$W$ provide a solution of the Warner problem. More precisely, it
was argued there that the category of topological B-branes in a
Landau-Ginzburg model is equivalent to the category $\mf(W)$
we have described in the previous subsection. The extent to which
$\mf(W)$ also describes ``physical'' branes in the untwisted 
Landau-Ginzburg model will be the subject of the present paper.

In the LG application, the space $M$ is the Chan-Paton space of a 
(target spacetime filling) DDbar-system, with equal number of 
branes and antibranes, and $f$ and $g$ describe a tachyon 
configuration. The shift functor $[1]$ is nothing but the 
exchange of branes with antibranes. The matrix $Q$ is part of 
the BRST charge, and the matrix factorization equation $Q^2=W$ 
is the condition that the tachyon configuration be BRST invariant
(or preserve $\caln=2$ supersymmetry in the untwisted model). 
Open string states between two such brane systems are given by 
the cohomology of $D$, \ie, are elements of $H^*(Q,Q')$. ($H^0$ 
being referred to as bosonic, and $H^1$ as fermionic.) 
The cone \eqref{cone} describes the formation of
a ``topological bound sate'' between two such configurations.
Finally, \eqref{trivial} simply corresponds to the addition of 
a brane-antibrane pair which is canceled by an identical
tachyon.

For other recent work on matrix factorizations in their relation
to D-branes in Landau-Ginzburg models, see \cite{kali2,kali3,calin27,
aad,hor2,hll,rutgers,hl,hll2,khro,bhlw,gsv,howa}.

\section{Graded matrix factorizations}
\label{graded}

By construction so far, our D-brane category $\mf(W)$ is 
$\zet_2$-graded. In particular, the shift functor squares to the
identity. On the other hand, the prime example of a
triangulated category, namely the derived category of coherent
sheaves on an algebraic variety $\DD(X)$, is $\zet$-graded (and
has shifts by arbitrary integers). As pointed out in \cite{howa}, 
there is a simple way to improve $\mf(W)$ to a $\zet$-graded 
category in the special case that $W$ is quasi-homogeneous. 

\subsection{R-symmetry}

$W$ being quasi-homogeneous is the condition that there exists
an assignment of degrees to the variables $x_i$ such that
$W$ has definite degree. In the physical model, this grading is 
worldsheet R-charge, and $W$ having R-charge $2$ is the 
conventional normalization. Thus, we assume that there exist 
R-charges $q_i\in\rationals$ such that
\begin{equation}
W(\ee^{\ii\lambda q_i} x_i) = \ee^{2\ii\lambda} W(x_i)
\qquad \text{ for all $\lambda\in\reals$}
\eqlabel{bulk}
\end{equation}
One can think of R-charge as a $\U(1)$ (or $\complex^\times$)
action on the space of polynomials with respect to which 
$W$ is equivariant. The $\U(1)$ action closes for $\lambda=
\pi H$, where $H$ is the smallest integer such that $H q_i\in2\zet$
for all $i$. 

When considering matrix factorizations of $W$, it is natural to 
require that this $\U(1)$ action can be extended to $(M,\sigma,Q)$.
This condition that the boundary interactions preserve the $\U(1)$ 
R-symmetry is a necessary condition for the existence of a conformal
IR fixed point. We will call such matrix factorizations {\it
quasi-homogeneous}. For compatibility with \eqref{matfac}, we must 
require that $Q$ has R-charge $1$. We will, at first, assume that
this $\U(1)$ acts on $M$ as an $\calr=\complex[x_1,\ldots,x_r]$-module
(instead of as a $\complex$-vectorspace). We will, however, assume
that the action is even, \ie, commutes with $\sigma$. Explicitly, 
we assume that there exists a map
\begin{equation}
\rho: \reals \to \GL^+(2N,\calr)= \GL(N,\calr) \times \GL(N,\calr)\,,
\eqlabel{rho}
\end{equation}
such that
\begin{align}
\rho(0,x_i) =\rho(\pi H,x_i) &=\id_{2N\times 2N} 
\eqlabel{closure} \\
\rho(\lambda,x_i) Q(\ee^{\ii\lambda q_i} x_i) 
&= \ee^{\ii\lambda} Q(x_i)\rho(\lambda,x_i)  \,.
\eqlabel{boundary} 
\intertext{Note that this implies the slightly non-standard group law}
\rho(\lambda,x_i) \rho(\lambda',\ee^{\ii\lambda q_i} x_i) &= 
\rho(\lambda+\lambda',x_i)  \,.
\eqlabel{group} 
\end{align}
In \eqref{rho}, $\GL(N,\calr)$ is the group of invertible $N\times N$
matrices with polynomial entries. Under gauge transformations,
$Q(x_i)\to U(x_i) Q(x_i) U(x_i)^{-1}$ with $U\in \GL^+(2N,\calr)$, 
$\rho$ transforms as
\begin{equation}
\rho^U(\lambda,x_i) = U(x_i) \rho(\lambda,x_i) 
U(\ee^{\ii\lambda q_i} x_i)^{-1} \,.
\eqlabel{garho}
\end{equation}
Note that if we can find a gauge transformation such that $\rho^U$ 
is diagonal, then by \eqref{closure}, $\rho^U$ must be independent 
of the $x_i$. Hence $\rho^U$ is an ordinary $\U(1)$ representation 
on $M$ as a $\complex$-vectorspace. On general grounds, one expects 
that one can always find such gauge transformation that makes $\rho$ 
diagonal. We will assume that this is true. But, as will become clear 
later, we do not want to exclude altogether gauge transformations of 
nonzero degree which might make $\rho$ non-diagonal (and 
$x_i$-dependent).

\subsection{Gradability is a topological condition}

Consider the vector field generating the $\U(1)$ action \eqref{rho}
\begin{equation}
R(\lambda,x_i) = -\ii \del_\lambda \rho(\lambda,x_i) 
\rho(\lambda,x_i)^{-1} \,.
\end{equation}
In general, this will depend on $\lambda$, but it is easy to see
that $R(\lambda,x_i)$ is actually determined for all $\lambda$
by \eqref{group} and $R(0,x_i)$. At $\lambda=0$, the condition
\eqref{boundary} becomes
\begin{equation}
EQ + [R,Q] = Q \,,
\eqlabel{cohomological}
\end{equation}
where
\begin{equation}
E = \sum_i q_i x_i \frac{\del}{\del x_i}
\end{equation}
is the ``Euler vector field''. Note that $W$ being quasi-homogeneous 
means $EW =2W$, and therefore, if $Q^2=W$,
\begin{equation}
\{Q,EQ - Q\} = 0 \,,
\end{equation} 
where $\{\cdot,\cdot\}$ is the anticommutator. In other words,
$EQ-Q$ defines a class in $H^1(Q)$. The existence of $R$ is the
statement that this class is trivial. 

The quasi-homogeneity condition on matrix factorizations is
therefore a topological condition that is roughly analogous,
by mirror symmetry, to the vanishing of the Maslov class
of Lagrangian cycles. Recall \cite{seidel} that the vanishing 
of the Maslov class ensures that Floer cohomology can be 
$\zet$-graded. Here, requiring \eqref{cohomological} will
immediately only give a $\rationals$-grading, which commutes
with the $\zet_2$ grading. In subsection \ref{orbifold}, we will 
combine the two gradings into a single $\zet$-grading.

Actually, requiring the infinitesimal version \eqref{cohomological}
is somewhat weaker than the integrated version \eqref{closure},
\eqref{boundary}, because it does not guarantee that $R$ 
generates a compact $\U(1)$ action. Equivalently, we might
not be able to diagonalize $R$ by a gauge transformation.
Deferring a discussion of this point to subsection \ref{Runique}, 
let us assume that $R$ is diagonalized (and hence its entries are
in $\rationals$). We then obtain an induced (diagonalizable)
$\U(1)$-action on the morphism spaces in the 
dg-category $\free(W)$. $\rationals$-homogeneous elements of 
$\Hom_{\free(W)}(Q,Q')$ satisfy
\begin{equation}
E\Phi + R'\Phi - \Phi R = q_\Phi \Phi
\eqlabel{morgrad}
\end{equation}
By \eqref{cohomological}, this descends to a $\rationals$-grading of 
$D$-cohomology, and hence, of $\mf(W)$. To avoid confusion, we will
use $\HH^*(Q,Q')=\oplus_{q\in\rationals}\HH^q(Q,Q')$ to denote this 
$\rationals$-graded cohomology, and also use the split
\begin{equation}
\HH^{q}(Q,Q') = \HH^{q,0}(Q,Q')\oplus \HH^{q,1}(Q,Q')
\end{equation}
into $\zet_2$ even and odd pieces.

\subsection{Serre duality}

If the boundary tachyon configuration described by $Q$ and $Q'$ flows 
to a conformal theory in the IR, one expects the spectrum of Ramond
ground states to be charge conjugation symmetric. As usual \cite{chiral},
by spectral flow, this means for the chiral primaries, which are given 
by $D$-cohomology
\begin{equation}
\begin{split}
H^*(Q,Q') &= H^{*+r}(Q',Q) \\
\HH^q(Q,Q') &= \HH^{\hat c-q}(Q',Q)
\end{split}
\eqlabel{serre}
\end{equation}
for the $\zet_2$ and $\rationals$-graded cohomologies, respectively.
Here $\hat c=\sum_{i=1}^r(1-q_i)$ is the central charge of the bulk
CFT associated with $W$. In mathematical terms, \eqref{serre} expresses 
``Serre duality'' for the category $\mf(W)$, with trivial Serre functor 
given purely by a shift in rational degree by $\hat c$, and reversal
of $\zet_2$ degree if the number of variables is odd.

Serre duality is equivalent to non-degeneracy of the boundary topological
metric, which was computed in \cite{kali2,hl}. If $\Phi\in\Hom_{\mf(W)}
(Q,Q')$ and $\Psi\in\Hom_{\mf(W)}(Q',Q)$, this Serre pairing is given by
\begin{equation}
\langle \Psi\Phi \rangle = 
\oint \frac{\str_M\left[(\del Q)^{\wedge r} \Psi \Phi\right]}
{\del_1 W\cdots\del_r W}
\eqlabel{pairing}
\end{equation}
where the integral is a multi-dimensional residue. It is easy to see that 
this pairing has $\rationals$-degree $\hat c$, \ie, $\langle\Psi\Phi
\rangle=0$ unless $q_\Phi+q_\Psi=\hat c$. It also has $\zet_2$ grading 
given by $r$, the number of variables in the model. Thus, proving 
non-degeneracy of \eqref{pairing} is equivalent to \eqref{serre}. It 
would be interesting to show this.

\subsection{Ambiguities of $R$}
\label{Runique}

As we have mentioned, the condition \eqref{cohomological} does
not guarantee that $R$ generates a compact $\U(1)$ action that
closes for $\lambda=\pi H$. On the other hand, it determines 
$R$ only up to an even matrix that commutes with $Q$, \ie, a 
representative of $H^0(Q)$. I am not aware of any example in 
which \eqref{cohomological} has a solution, but no solution 
which does not generate a compact $\U(1)$ action, or which is
not diagonalizable.

For example, if all entries of $Q=\left(\begin{smallmatrix}
0 & f\\g&0\end{smallmatrix}\right)$ are in fact homogeneous 
polynomials, then one expects that \eqref{cohomological}
generically has a solution $R={\rm diag}(R_1,\ldots,R_{2N})$ 
which is diagonal. Indeed, denoting polynomial degree by $\deg$,
equation \eqref{cohomological} becomes
\begin{equation}
\begin{split}
R_j-R_{k+N} &= 1 - \deg(f_{jk})  \\
R_{k+N}- R_j &= 1 - \deg(g_{kj}) 
\end{split}
\qquad \text{for $j,k=1,\ldots,N$}\,, 
\eqlabel{over}
\end{equation}
which is a system of $2N^2$ equations for $2N$ unknowns.
The non-trivial relations on the left hand side of 
\eqref{over} are given by permutations $\pi\in\Sigma_N$ on 
$N$ indices,
\begin{equation}
\sum_{j} \left(R_j - R_{\pi(j)+N} \right)
\end{equation}
being independent of $\pi$. On the right hand side, these
relations become
\begin{equation}
N - \sum_j \deg(f_{j\pi(j)}) \,.
\eqlabel{rhsrels}
\end{equation}
On the other hand, on taking determinant of \eqref{mafa}, we
see that
\begin{equation} 
\det(f) \det(g) = W^N \,,
\end{equation}
which, assuming that $W$ is irreducible, implies $\det(f)=W^k$
for some $0\le k\le N$. Since $\sum \deg(f_{j\pi(j)})$ is
the degree of a summand of $\det(f)$, we see that if there are
no exceptional cancellations, \eqref{rhsrels} is independent of 
$\pi$. Similarly, $g f = W$ generically implies $\deg(f_{jk})+
\deg(g_{kj})=2$. 

Thus if all entries of $Q$ are homogeneous, we expect that there 
is a diagonal solution of \eqref{cohomological} (this is true in 
all examples I have studied). It is easy to see that the
converse is also true: If $R$ is diagonal, then all entries of 
$Q$ must be homogeneous. (But there are factorizations that are
not quasi-homogeneous, see subsection \ref{nonhomo}!)

We will generally assume that there is a solution of 
\eqref{cohomological} that is diagonalizable, keeping in 
mind that this assumption can conceivably fail at singular
loci in the moduli space of matrix factorizations.
Let us then analyze the ambiguities of $R$. 

The proposal for fixing the ambiguity of $R$ is motivated by 
the examples of section \ref{examples} and the general considerations 
of section \ref{stabcon}. \footnote{It is also reminiscent 
of the ``$a$-maximization'' procedure used to find the R-charge
of $\caln=1$ superconformal gauge theories in four dimensions
\cite{inwe}.} The essential idea is that $R$ defines a character 
on the gauge group of similarity transformations. Infinitesimally, 
such gauge transformations are given by even endomorphisms of $M$ 
as a $\calr$-module, \ie, block-diagonal matrices $V\in{\rm 
Mat}^+(2N\times 2N,\calr)$, with $\tr V\in \complex$. They act 
on $Q$ by $\delta Q =[V,Q]$, and the character induced by $R$ 
is given by
\begin{equation}
\chi_R(V)= \tr_M(R V) \,.
\end{equation}
The condition we would like to impose on $R$ is that this
character be trivial on the part of the gauge group acting 
trivially,
\begin{equation}
\tr(R V) = 0 \qquad\text{whenever}\qquad  [V,Q]=0 \,.
\eqlabel{condition}
\end{equation}
Note that under such infinitesimal gauge transformations,
$R$ transforms according to
\begin{equation}
\delta R = - EV - [R,V]
\eqlabel{Rtra}
\end{equation}
which leaves \eqref{cohomological} invariant to first order.
By all we have said, it might then seem natural to fix a 
diagonal $R$ and restrict to gauge transformation of degree
$0$, \ie, those which satisfy $EV+[R,V]=0$ leave $R$ invariant. 
As we will see in section \ref{examples}, however, this would 
be too restrictive, as we would not be able to describe 
brane-antibrane annihilation.

To fix the ambiguity, and impose $\eqref{condition}$, one may 
proceed as follows. Start with a reference solution $R_0$, 
assumed to be diagonal. The ambiguities of \eqref{cohomological}, 
which are parametrized by even cycles of $Q$, can be decomposed
according to the degree with respect to $R_0$,
\begin{equation}
C^0(Q) = \oplus_{q} \CC^{q,0}(Q) 
\end{equation}
where
\begin{equation}
\CC^{q,0} = \left\{ V\in {\rm Mat}(2N\times 2N,\calr) ;
[Q,V]=0\,, [\sigma,V]=0\,,
EV+[R_0,V] =q V \right\} \,.
\end{equation}

To see what can happen if we modify $R_0$ by an element of 
$C^0(Q)$, it is instructive to consider the following example.
Let
\begin{equation}
R_0 = \begin{pmatrix} a & 0 \\0 & b \end{pmatrix}
\end{equation}
with $a$ and $b$ rational. Then (we are neglecting $Q$ and
$\sigma$ in this discussion---$R_0$ and $V$ could be 
submatrices in a larger problem),
\begin{equation}
V = \begin{pmatrix} 0 & x \\ 0 & 0 \end{pmatrix}
\end{equation}
is of (total) degree $q(V)=\deg(x)+a-b$ with respect to 
$R_0$. Clearly, $R_0$ generates a compact $\U(1)$ by 
$\rho_0(\lambda)=\ee^{\ii\lambda R_0}$, but $R_0+V$
not necessarily so. Indeed, it is easy to see that the 
solution of \eqref{group} generated by $R_0+V$ at 
$\lambda=0$ is
\begin{equation}
\rho(\lambda,x) = 
\begin{pmatrix}
\ee^{\ii\lambda a} & x \frac{\ee^{\ii \lambda (a+\deg(x))}
-\ee^{\ii \lambda b}}{a+\deg(x)-b} \\
0 & \ee^{\ii\lambda b}  
\end{pmatrix} \,, 
\end{equation}
which for $b=a+\deg(x)$ goes over into
\begin{equation}
\rho(\lambda,x)=
\begin{pmatrix}
\ee^{\ii\lambda a} & \ii x \lambda\ee^{\ii\lambda b} \\
0 & \ee^{\ii\lambda b}
\end{pmatrix} \,.
\end{equation}
We see that as long as $q(V)=\deg(x)+a-b\neq 0$, $R=R_0+V$
generates a compact $\U(1)$ and can be diagonalized.
This fails when $q(V)=0$.

Thus, by rediagonalizing $R$ if necessary, we can neglect 
the modifications of $R_0$ by elements of $\CC^{q,0}(Q)$
for $q\neq 0$. And clearly, $\chi_{R_0}$ vanishes automatically
on $V\in\CC^{q,0}(Q)$, because such $V$ doesn't have diagonal 
entries.

What about the ambiguities parametrized by $\CC^{0,0}(Q)$?
As we have just seen, we cannot add $V$'s without diagonal
entries. Among those with diagonal entries, we choose a
maximal commuting subalgebra, with basis $\{V_i\}_{i=1,
\ldots,s}$, and impose \eqref{condition} on the ansatz
\begin{equation}
R = R_0 + \sum_{i=1}^s a_i V_i \,.
\eqlabel{ansatz}
\end{equation}
In all examples I have studied, this procedure leads to an 
unambiguously determined $R$ which is diagonalizable and 
satisfies \eqref{condition}.

\subsection{Cone construction}
\label{conegrad}

We next show that the grading we just introduced is compatible with 
the triangulated structure. In particular, we show that the cone
\eqref{cone} over a map $T[1]\in H^0(Q_1,Q_2[1])$ between two 
quasi-homogeneous matrix factorizations is again 
quasi-homogeneous.

Indeed, 
\begin{equation}
\begin{split}
EQ-Q &= \begin{pmatrix} E Q_1 -Q_1 & 0 \\ E T-T & EQ_2-Q_2
\end{pmatrix} \\
&= \begin{pmatrix} [Q_1,R_1] & 0 \\
(q_T-1)T-R_2T+TR_1 & [Q_2,R_2] 
\end{pmatrix}  \\
&= \left[Q,
\begin{pmatrix} 
R_1+ (q_T-1) \id_1 & 0 \\ 
0 & R_2 
\end{pmatrix}
\right]
\end{split}
\eqlabel{standard}
\end{equation}
is exact. More properly, we could choose
\begin{equation}
R = \begin{pmatrix} R_1 +  (q_T-1)\frac{N_2}{N_1+N_2}\id_1 & 0 \\
0 & R_2 -(q_T-1) \frac{N_1}{N_1+N_2}\id_2\end{pmatrix}
\eqlabel{choose}
\end{equation}
so as to satisfy $\tr(R)=0$ as well as $EQ-Q=[Q,R]$. But in
the generic case, $C^0(Q)$ will contain more elements than just the 
identity so that we will not satisfy \eqref{condition} in general.

\subsection{Orbifolding and the phase of matrix factorizations}
\label{orbifold}

Recall that the Calabi-Yau/Landau-Ginzburg correspondence relates
Calabi-Yau manifolds given as complete intersections in toric
varieties to Landau-Ginzburg orbifold models \cite{glsm}. In the 
simplest case, the Calabi-Yau is a hypersurface $X$ given as the
vanishing locus of a polynomial $P$ of total degree $H$ in
weighted projective space $\projective_{w_1,\ldots,w_r}^{r-1}$
such that $\sum_{i=1}^r w_i = H$. Such an $X$ corresponds, via 
CY/LG correspondence, to the Landau-Ginzburg orbifold model with 
superpotential $W=P$ and orbifold group $\Gamma=\zet_H$, with 
central charge $\hat c=r-2$. The $x_i$ have R-charge 
$q_i=2w_i/H$ and $\Gamma$ is generated by $x_i\mapsto 
\omega_i x_i$ with $\omega_i=\ee^{\ii\pi q_i}$. The R-charges
of the invariant part of the bulk chiral ring 
$\calj=\left(\calr/\del W\right)^\Gamma$ are then all even
integers.

Now let $Q$ be a quasi-homogeneous matrix factorization of $W$
with R-matrix $R$ uniquely determined as in subsection 
\eqref{Runique}, and diagonal. Obviously, we would like to extend 
the $\Gamma$ action to $Q$, and we require that it commutes with
the rational and the $\zet_2$-grading. In other words, we are 
looking for a representation of $\Gamma$ on (the associated 
$\zet_2$-graded $\calr$-module of CP factors) $M$ such that
\begin{equation}
\gamma Q(\omega_i x_i) \gamma^{-1} = Q(x_i) \,.
\eqlabel{equi}
\end{equation}
It is easy to see that such a representation must be related to
$R$ in a simple manner. Indeed, we see that
\begin{equation}
\tilde \gamma = \sigma \ee^{-\ii\pi R} \gamma
\eqlabel{orbi}
\end{equation}
commutes with $Q$. It is no restriction to assume that
it is diagonal. If $Q$ is reduced (\ie, contains not entries
with a constant term), then all diagonal degree 
$0$ elements of $C^0(Q)$ are actually non-trivial in 
$\HH^{0,0}(Q)$. We conclude that if $Q$ is reduced and 
irreducible (\ie, $\HH^{0,0}(Q)$ is one-dimensional), then 
$\tilde\gamma$ is a multiple of the identity, $\tilde\gamma=
\ee^{\ii\pi\varphi}$. In other words, we find
\begin{equation}
\gamma = \sigma \ee^{\ii\pi R} \ee^{-\ii\pi\varphi}
\eqlabel{fix}
\end{equation}
Imposing $\gamma^H=1$ fixes $\varphi\in\reals\bmod 2/H$. 
Lifting to $\varphi\in\reals\bmod 2$ gives $H$ different
equivariant factorizations for each factorization of $W$.
(To be sure, if $H$ is even, these correspond to $H/2$ branes
together with their antibranes.)

A $\Gamma$-action on the objects induces an action on the morphism 
spaces $\HH^*(Q,Q')$, and we can project onto invariant morphisms
by requiring
\begin{equation}
\gamma' \Phi(\omega_ix_i) \gamma^{-1} = \Phi
\eqlabel{morproj}
\end{equation}
By combining the definitions, it is easy to see that invariant 
morphisms satisfy the condition
\begin{equation}
\ee^{\ii \pi q_\Phi} (-1)^\Phi \ee^{\pi\ii(\varphi-\varphi')}
=1
\end{equation}
In other words, $q_\Phi=\varphi'-\varphi+n$, where $n$
has the same parity as $\Phi$. This constraint on the $\U(1)$
charges is the same as \eqref{qmod}, and leads to the identification 
of $\varphi$ as the phase of the matrix factorization.

Let us then define the category $\MF(W)$, in which objects
are quasi-homogeneous matrix factorizations $Q$ together with a 
lift of the phase $\varphi$ to a real ``grade'', and morphism 
spaces are
\begin{equation}
\Hom^n\bigl((Q,\varphi),(Q',\varphi')\bigr) = 
\HH^{q=n+\varphi'-\varphi}(Q,Q')
\eqlabel{zgraddef}
\end{equation}
This is the promised $\zet$-graded category of matrix factorizations.
Note in particular that the shift functor, which, because of 
\eqref{orbi}, must be accompanied by $\varphi\to \varphi+1$, does 
not square to the identity in $\MF(W)$. 

\subsection{Conjecture}

The general decoupling statements of \cite{douglas}, the result
that B-branes are described at large volume by the derived
category of coherent sheaves, together with the assumption
that all topological B-branes of the Landau-Ginzburg model
have a description using matrix factorizations, naturally lead 
to the statement that---in appropriate cases---there should be 
an equivalence of categories
\begin{equation}
\MF(W) \cong \DD(X) \,,
\eqlabel{conj}
\end{equation}
where $\MF(W)$ is the category of quasi-homogeneous $\Gamma$-equivariant
matrix factorizations of $W$ and $\DD(X)$ is the derived category of
coherent sheaves on the Calabi-Yau manifold $X$ related to $W/\Gamma$
by Witten's gauged linear sigma model construction \cite{glsm}.
Cases in which one expects such a correspondence include
those GLSM's in which both large volume and Landau-Ginzburg
points exist and are unique, such as the quintic in 
$\projective^4$, or hypersurfaces in weighted projective
spaces.

I hope that such a correspondence appears well-motivated
from the physics point of view. It has essentially already
been stated by Ashok, Dell'Aquila and Diaconescu in \cite{aad} 
(for the quintic case and without the homogeneity condition). 
I should, however, add that the correspondence is somewhat different 
from existing (and mathematically proven!) equivalences between 
categories of matrix factorizations and other structures. 
Besides Eisenbud's canonical correspondence \cite{eisen} with maximal 
Cohen-Macaulay modules, there is also an equivalence between matrix 
factorizations and a so-called ``triangulated category of singularities'' 
which was proven by Orlov \cite{orlov}. Moreover, there is the classical 
correspondence of Grothendieck and Serre between graded modules 
over graded rings and vector bundles over the associated projective
variety. This correspondence was exploited by Laza, Pfister and
Popescu in \cite{laza} for the case of the elliptic curve.
If \eqref{conj} is true, it is likely that the equivalence
favored by physics is different from those just mentioned. For 
the elliptic curve, for instance, the methods of \cite{laza} on 
the one hand and \cite{bdlr,bhlw} on the other hand yield quite 
different bundles corresponding to some given matrix factorization.

\subsection{Landau-Ginzburg monodromy} 

We can make one further check that our conjecture makes 
sense. The Landau-Ginzburg point $p_{\rm LG}$ is an orbifold
point in the K\"ahler moduli space $\calm_k$. The action 
of the monodromy around $p_{\rm LG}$ acts on matrix factorizations
in $\MF(W)$ simply by rotating the choice of lift of 
$\varphi$ in \eqref{fix},
\begin{equation}
\varphi\to \varphi+2/H \,.
\end{equation}
As a consequence, the $H$-th power of the Landau-Ginzburg
monodromy operator acts by $\varphi\to\varphi+2$. This does
nothing on the physical brane associated with $Q$, but is 
a shift by $2$ in the triangulated category. This solves
a problem posed in \cite{aspreview}, in which the $5$-th
power of the Landau-Ginzburg monodromy on the quintic
Calabi-Yau was computed and found to correspond to a
shift by $2$ on the derived category. We can simply confirm 
this result using matrix factorizations, and in fact extend 
it to all Calabi-Yau manifolds with a Landau-Ginzburg 
description.

\section{RR charges and index theorem}
\label{index}

If matrix factorizations represent D-branes in string 
theory, they must carry Ramond-Ramond (RR) charge. This 
charge takes value in the dual of the appropriate space 
$\calh_{\rm RR}^{\rm B}$ of closed string RR ground states. 
Because of the boundary condition on the worldsheet $\U(1)$ 
current, B-branes couple to those RR ground states with opposite 
left and right-moving R-charge, $q_L=-q_R$. The purpose of this
section is to determine these RR charges of matrix 
factorizations. We first describe $\calh_{\rm RR}^{\rm B}$.

The space of Ramond-Ramond ground states in Landau-Ginzburg
orbifolds and their left-right R-charges was computed in 
\cite{lgorb}. For simplicity, we will restrict here to a 
cyclic orbifold group $\Gamma=\zet_H$, as well as to integer 
central charge (we mostly have in mind, of course, 
$\hat c=3$). The generalization of at least some of the 
formulas to the more general case should be obvious. 
In general, those RR ground states with $q_L\neq q_R$ arise 
purely from the twisted sector, and if $\hat c$ is integer,
the $\zet_H$ projection on twisted sectors implies that the 
RR ground states have $q_L\equiv q_R\equiv \hat c/2\bmod\zet$.

Consider the $l$-th twisted sector, and divide the field
variables of the LG model into two classes, according to
whether $l q_i\in2\zet$ or $l q_i\notin 2\zet$. Those 
fields, $\{x_i^t\}$, with $l q_i\notin 2\zet$ satisfy twisted 
boundary condition in this sector, and must be set to zero in 
the semi-classical analysis used to determine the RR ground 
states. The contribution of those fields to the R-charges 
is\footnote{What we have called $q_i$ is the sum of left-
and right-moving charges of the variables $x_i$ in the 
normalization of \cite{lgorb}.}
\begin{equation}
q_L^t = - q_R^t = \sum_{lq_i\notin\zet}
\Bigl(l\frac{q_i}2-\Bigl[l\frac{q_i}2\Bigr]-\frac12\Bigr)
\end{equation}
On the other hand, those fields, $\{x_i^u\}_{i=1,\ldots,r_l}$, 
with $lq_i\in 2\zet$ satisfy untwisted boundary conditions in 
the $l$-th twisted sector. Their quantization leads to a spectrum 
of RR ground states which is that corresponding to the effective 
potential $W_l(x_i^u)=W(x_i^u,x_i^t=0)$. In particular, they 
contribute, $q_L^u=q_R^u$, equal amounts to left and right charge.

What is important for us, the ground states with $q_L=q_L^u+
q_L^t=-q_R=q_R^u+q_R^t$ from the $l$-th twisted sector correspond
precisely to the neutral ground states of the effective potential 
$W_l(x_i^u)$ obtained by setting those fields with $l q_i\notin 
2\zet$ to zero. These ground states have $q_L\equiv q_R\equiv\hat c/2
\bmod\zet$ if the number, $r_l$, of fields with $lq_i\in2\zet$ is even.
A basis of these ground states can be labeled as $|l;\alpha\rangle$,
where $l$ ranges between $0$ and $H-1$, and $\alpha$ ranges over
a basis $\phi_l^\alpha=(x^u)^{\alpha}=\prod_{i=1}^{r_l} 
(x_i^u)^{\alpha_i}$ of the subspace, $\calj_l^0$, of the untwisted 
chiral ring $\calj_l=\complex[x_i^u]/\del W_l$ with R-charge 
$q_L^u=q_R^u= \sum_{lq_i\in 2\zet} \alpha_i q_i/2=\hat c^u/2$. Here, 
$\hat c^u= \sum_{lq_i\in2\zet}(1-q_i)$ is the central charge 
corresponding to $W_l$. The states $|l;\alpha\rangle$ can be thought 
of as being obtained by acting with $\phi^l_\alpha$ on the unique 
state $|l;0\rangle$, which has R-charge $-\hat c^u/2$.

Now by definition, the RR charge is the correlation function on the
disk with the RR ground state inserted in the bulk. We propose that
for a matrix factorization $Q\in\MF(W)$, this is given by
\begin{equation}
\begin{split}
\ch(Q)&: \calh_{\rm RR}^{\rm B} \to \complex \\[.2cm]
\ch(Q)(|l;\alpha\rangle) 
&= \langle l;\alpha|Q\rangle_{\rm disk} \\
&= \frac{1}{r_l!}\res_{W_l} \bigl( \phi_l^\alpha\; \str\bigl[\gamma^l 
(\del Q_l)^{\wedge r_l}\bigr] \bigr) \\
&= \frac {1}{r_l!}\oint \frac{\phi_l^\alpha\;\str\left[\gamma^l 
(\del Q_l)^{\wedge r_l} \right]}{\del_1 W_l\cdots\del_{r_l} W_l}
\end{split}
\eqlabel{chern}
\end{equation}
where $\gamma$ is the representation of the generator of
$\zet_H$ on the matrix factorization, and $\str(\,\cdot)\,)=
\tr_M(\sigma\,\cdot\,)$ is the supertrace over the $\zet_2$-graded 
module $M$. The residue is the same as the one appearing in the Serre 
pairing \eqref{pairing}. It is normalized \cite{toplg} such that the 
determinant of the Hessian of the superpotential $W_l$ has residue 
equal to the dimension of the chiral ring,
\begin{equation}
\res_l(\det \del_i\del_j W_l)
=\oint \frac{\det \del_i\del_j W_l}{\del_1 W_l\cdots \del_{r_l} 
W_l} = \dim\calj_l = \mu_l =  \prod_{lq_i\in2\zet} \frac{2-q_i}{q_i} \,.
\end{equation}
Moreover, in \eqref{chern}, $Q_l(x_i^u)=Q(x_i^u,x_i^t=0)$ is the 
restriction of $Q$ to the untwisted fields in the $l$-th sector. It 
satisfies $Q_l^2=W_l$.

Formula \eqref{chern} makes sense since by \eqref{equi},
\begin{equation}
\gamma^l Q^l (x_i^u)  = Q^l(x_i^u) \gamma^l \,,
\end{equation}
so $\gamma^l$ represents a cohomology class of the matrix factorization
$Q_l$, and \eqref{chern} computes the disk correlation function 
\cite{kali2,hl} of $\gamma^l\phi^\alpha_l$ in this model.
Moreover, since $\res_l$ has $\rationals$ degree $\hat c^u$
and $\zet_2$ degree $r_l$, we see that \eqref{chern} would vanish if
$r_l$ were odd or if we tried to insert an element of $\calj_l$ 
with charge not equal to $\hat c^u$.

In those twisted sectors with $r_l=0$, \ie, $lq_i\notin 2\zet$ for all
$i$, \eqref{chern} reduces to
\begin{equation}
\ch(Q)(|l;0\rangle) = \str \gamma^l \,.
\end{equation}

One can check that \eqref{chern} gives the correct value for the RR charges 
in those cases where an alternate computation exists, namely minimal models 
and their tensor products. Also, we see immediately that $\ch(Q[1])=-\ch(Q)$. 
The main evidence, however, that \eqref{chern} is the correct expression 
for the RR charge is the index theorem for matrix factorizations,
\ie, the fact that the Witten index for open strings between two matrix
factorization $Q$ and $Q'$ can be computed via $\ch(Q)$ and $\ch(Q')$ as
\begin{equation}
\begin{split}
\tr (-1)^F &=  
\sum_{n\in\zet} (-1)^n \dim\Hom_{\MF(W)}^n(Q,Q')  \\
&=\langle\ch(Q'),\ch(Q)\rangle
\end{split}
\eqlabel{indtheo}
\end{equation}
where $(-1)^F$ is the $\zet_2$ grading (fermion number) of matrix
factorizations. The Chern pairing is given by
\begin{equation}
\langle\ch(Q'),\ch(Q)\rangle
= \frac1H\sum_{l=0}^{H-1} \sum_{\alpha,\beta}
\ch(Q')(|l;\alpha\rangle)
\frac{1}{\prod_{lq_i\notin 2\zet} (1-\omega_i^l)}
\eta_l^{\alpha\beta}
\ch(Q)(|l;\beta\rangle)^*
\eqlabel{chpa}
\end{equation}
For fixed $l$, $\sum_{\alpha,\beta}$ is a sum over the chosen basis of 
$\calj_l^0$ of elements of the chiral ring $\calj_l$ with charge 
$\hat c^u/2$, and $\eta_l^{\alpha\beta}$ is the inverse of the closed string 
topological metric in this sector,
\begin{equation}
\eta^l_{\alpha\beta} = \res_l\bigl(\phi_l^\alpha\phi_l^\beta\bigr) \,.
\eqlabel{topmet}
\end{equation}

We will now prove \eqref{indtheo} in the case that $r_l=0$ in all twisted 
sectors. The index of interest is the equivariant index of the operator
$D$ acting as in \eqref{superco} on the complex given by the morphism
space in $\free(W)$, \ie,
\begin{equation}
\tr(-1)^F = \frac 1H\sum_{l=0}^{H-1} \tr(-1)^F \tilde\gamma^l \,,
\end{equation}
where $\tilde\gamma$ is the action of the generator of $\zet_H$
on the cohomology spaces. We can regularize the computation 
of $\tr(-1)^F \tilde\gamma = \lim_{t\to 1} Z_l(t)$ by using
the $\rationals$-grading by $\U(1)$ charge
\begin{equation}
Z_l(t) = \tr(-1)^F t^{q} \tilde\gamma^l \,.
\end{equation}
(More precisely, we should use an appropriate covering of this
$\U(1)$ to make the charges integer.) By a standard argument, we can 
then replace the trace over the space of ground states by the trace over 
$\Hom_{\free(W)}(Q,Q')=\Hom_{\calr}(M,M')$, effectively reducing the 
computation to the setting $Q=Q'=0$. We decompose 
\begin{equation}
\Hom_\calr(M,M') =\bigoplus_{j,k=1}^{2N}
\bigoplus_{\alpha} \calv_{j,k,\alpha}
\end{equation}
into one-dimensional pieces indexed by matrix entries $(j,k)$ and
monomials $x^\alpha=\prod x_i^{\alpha_i}$ with multi-index
$\alpha=(\alpha_1,\ldots,\alpha_r)$. Note that the combination of 
fermion number and $\zet_H$-action restricts on $\calv_{j,k,\alpha}$
to
\begin{equation}
\left((-1)^F \tilde\gamma\right) |_{\calv_{j,k,\alpha}} =
\sigma'_j \gamma'_j \biggl(\prod_{i=1}^r \omega_i^{\alpha_i}
\biggr)
\sigma_k\gamma^{-1}_k
\end{equation}
where $\omega_i=\ee^{\pi\ii q_i}$, and we are using that both 
$\sigma$ and $\gamma$ are diagonal matrices. Therefore,
\begin{equation}
\begin{split}
Z_l(t) &= \sum_{j,k=1}^{2N} \sum_{\alpha}
\sigma'_j(\gamma'_j)^l\, t^{R'_j}\;
\biggl(\prod_{i=1}^r\omega_i^{l \alpha_i} t^{q_i\alpha_i} \biggr)\;
\sigma_k(\gamma_k)^{-l}\, t^{-R_k}  \\
&= \str\bigl((\gamma')^lt^{R'}\bigr) \frac 1{\prod_i(1-t^{q_i}\omega_i^l)}
\str\bigl(\gamma^{-l} t^{-R}\bigr) \,.
\end{split} 
\end{equation}
Since $\str(\id)=0$, and we are assuming that $lq_i\notin\zet$
for all other $l$ and $i$, we can smoothly take $t\to 1$, and obtain
\begin{equation}
\tr(-1)^F = \frac1H\sum_{l=1}^{H-1}
\str(\gamma')^l \frac1{\prod_i(1-\omega_i^l)}
\str\gamma^{-l} \,,
\end{equation}
as was to be shown.

To establish \eqref{indtheo} and \eqref{chpa} in general, one should
combine the proof we just gave with the formula
\begin{equation}
\begin{split}
\frac{1}{(r_l!)^2} 
\res_l \bigl( 
\str\bigl[(\del Q'_l)^{\wedge r_l}  \bigr] \;&
\str\bigl[(\del Q_l)^{\wedge r_l} \bigr] \bigr)  \\
=
\sum_{\alpha,\beta} 
\frac{1}{r_l!}
\res_l&\bigl(\phi_l^\alpha\;
\str\bigl[(\del Q'_l)^{\wedge r_l}\bigr]\bigr)
\;
\eta_l^{\alpha\beta}
\;
\frac{1}{r_l!}
\res_l\bigl(\phi_l^\beta\;
\str\bigl[(\del Q_l)^{\wedge r_l}\bigr]\bigr) \,.
\end{split}
\eqlabel{cardy}
\end{equation}
This formula expresses the factorization rule for the topological 
annulus correlator \cite{kali2} with no boundary insertions via 
two disk amplitudes and (the inverse of) the closed string topological 
metric $\eta^l_{\alpha,\beta}$ \eqref{topmet} given by the sphere amplitude 
\cite{toplg}. (Note that in \eqref{cardy}, the sum over $\alpha,\beta$ 
can be extended to the full chiral ring $\calj_l$ because the disk 
correlators vanish outside of $\calj_l^0$.) In the general axioms of 
open-closed topological field theory \cite{lazastructure,mose}, this 
factorization is known as the ``Cardy condition''. By the same axioms, 
the annulus correlator \eqref{cardy} computes the open string Witten 
index $\tr(-1)^F$ between the matrix factorizations $Q_l$, $Q'_l$ in 
the untwisted Landau-Ginzburg model corresponding to $W_l$. 
I have checked the equality of the two sides of \eqref{cardy}, 
and that they compute the open string Witten index, in all examples
I know, but I do not know a proof based directly on the residue
formula. 

We close this section with a few comments.

Firstly, we note that there is an obvious analogy between \eqref{indtheo}, 
\eqref{chpa} and the well-known Hirzebruch-Riemann-Roch theorem which 
computes the Witten index for open strings coupled to two vector bundles
$E$ and $F$ on the Calabi-Yau manifold $X$
\begin{equation}
\tr(-1)^F= \int \ch(E^*) \ch(F) \td(X) \,.
\end{equation} 
Our formula is simply the small volume version of this. In particular, 
the factor $\eta_l^{\alpha\beta}/\prod(1-\omega_i^l)$ can be viewed 
as the analog of the Todd class of $X$. From this perspective, 
the normalization in which the square-root of this factor is included in 
the charge might seem more natural.

Secondly, we return to the split of the Ramond ground states into those
from twisted sectors with $r_l=0$ and those from twisted sectors with
$r_l\neq 0$ and even. In the RCFT description of LG models as Gepner 
models \cite{gepner}, the ground states with $r_l\neq 0$ are not left-right
symmetric in each individual $\caln=2$ minimal model.\footnote{Geometrically, 
they correspond to non-toric blowups of $X$. For this reason, most
models that have been studied geometrically in any depth do not have
such states.} As a consequence, the BCFT constructions of boundary 
states in Gepner models \cite{resc} did not produce boundary 
states with charge under those RR ground states with 
$r_l\neq 0$, the only exception being related to 
the so-called fixed point resolution phenomenon discussed 
in \cite{brsc,fkllsw} (see also \cite{mms,kali2}). On the 
other hand, it is easy to find matrix factorizations for which 
charges with $r_l\neq 0$ do not vanish. (The two-variable 
factorizations of subsection \ref{d0brane}, when embedded 
in the appropriate Calabi-Yau model, provide useful examples.) 
It seems likely that matrix factorizations span the free part of 
K-theory that is expected from cohomology. What the Chern classes 
\eqref{chern} miss, of course, is the torsion part of the K-theory. 
Unorbifolded minimal models, for example, have K-theory that is purely 
torsion. One might expect that some of this will survive the orbifold 
procedure, conceivably in the twisted sectors with $r_l$ odd. It would 
be interesting to determine the full K-theory of these Landau-Ginzburg 
orbifolds and compare with their geometric computation. This would
be a zeroth order check of \eqref{conj}.

\section{A stability condition}
\label{stabcon}

In mathematical models of D-branes similar to the one we are 
studying, such as Lagrangian submanifolds of symplectic manifolds, 
holomorphic vector bundles on complex manifolds, or representations
of a quiver algebra, a stability condition is introduced with the 
purpose of identifying a subset of objects whose orbits under the
group of appropriate automorphisms fit together into ``nice'' 
moduli spaces. Often, the stable orbits admit a distinct (unique) 
representative at the zero of a ``moment map'' associated with 
the stability condition (for instance, the special condition for 
Lagrangians or the hermitian Yang-Mills equation for the connection 
on the holomorphic vector bundle). (See, \eg, chapter 38 of 
\cite{horietal} for a recount of these stories.)

In physics, the zeroes of the moment map are associated with the 
solution of the condition that the D-brane preserve supersymmetry 
in the uncompactified part of spacetime. Stability is the condition
that such a supersymmetric configuration can be reached by 
boundary renormalization group (RG) flow on the string worldsheet.
In the unstable (including semistable) case, the theory is
expected to split at singular points along RG flow into the direct
sum of several decoupled theories. The endpoint of the flow is the 
decomposition into the stable pieces.

For a general Landau-Ginzburg model (orbifolded or not, with
arbitrary central charge), the interpretation involving spacetime
supersymmetry is not necessarily available, and we will factor
it out accordingly. What remains is the unitarity constraint
\eqref{unitarity} and the assertion that if this condition is
satisfied, worldsheet RG flow should lead to a single unitary 
boundary CFT in the IR (\ie, a theory with a unique open string 
vacuum). This is a stability condition that can be imposed on 
the triangulated category of any quasi-homogeneous Landau-Ginzburg 
model. 

If the model has a geometric interpretation, then in view of the 
expected equivalence \eqref{conj}, this is a particular stability 
condition on $\DD(X)$. It is distinguished by the fact that
it arises only from data involving the unorbifolded model 
(or equivalently, the orbifolded model divided by the quantum 
symmetry). In the general framework of \cite{bridgeland}, the 
space of (numerical) stability conditions is locally modeled on 
the free part of the K-theory. As we have seen in section 
\ref{index}, most of the K-theory (all of it for an odd number of 
variables) appears during orbifolding. Therefore, the stability
condition in the Landau-Ginzburg model should be more rigid than
the ones on $\DD(X)$.

\subsection{A notion of stability}
\label{rstab}

As we have reviewed in section \ref{review}, the basic idea
underlying $\Pi$-stability is that open strings between 
physical branes should satisfy the unitarity constraint 
$0\le q\le\hat c$. It is hard, however, to impose such a constraint
directly on individual objects to determine whether they are
stable, essentially because this would involve an infinite
number of checks, and moreover because a stable object does
not only have strings satisfying \eqref{unitarity} ending on it.
Physically \cite{douglas}, one should not try to impose the 
condition \eqref{unitarity} on configurations described as 
(topological) boundstates containing both branes and antibranes.
One expects that in certain regimes \cite{douglas,asdo}, or even
at all points in the space of stability conditions \cite{bridgeland}, 
integrating out all canceling brane-antibrane pairs will reduce 
the problem to a stability condition on an Abelian category, which 
involves only a finite number of checks. Still, these discussions
leave open the question whether $\Pi$-stability is sufficient
or just a self-consistent ``bootstrap'' condition. Our point 
of view is that the Landau-Ginzburg model should posses an 
intrinsic (rigid!) stability condition that does not depend 
on what is going on in the rest of the moduli space. It is
this notion of stability that we are after.

In the Landau-Ginzburg context, ``integrating out brane-antibrane 
pairs'' simply corresponds to restricting to reduced matrix 
factorizations, \ie, those without scalar entries. It is not
unreasonable to expect, therefore, that by going to reduced 
matrix factorizations, one obtains the Abelian category of
interest for the discussion of \cite{bridgeland}. This Abelian 
category could be simply related to the category of 
Cohen-Macaulay module of subsection \ref{MCM}. In any case, we 
now make the following tentative definition.

Let $W(x_1,\ldots,x_r)$ be a quasi-homogeneous Landau-Ginzburg 
polynomial, $EW=2W$, where $E=\sum q_ix_i\del_i$. Let $Q$ be a 
reduced quasi-homogeneous matrix factorization of $W$. $Q$ is 
called R-semistable if in all triangles
\begin{equation}
\xymatrix@R=1cm@C=0.7cm{
& Q \ar[dr]^{S_1} & \\
\ar[ur]^{S_2} Q_2 && \ar@{-->}[ll]^{T} Q_1 
}
\end{equation}
in which $Q$ participates opposite to the fermionic morphism $T$,
we have
\begin{equation}
q_T \le 1 \;\; \Leftrightarrow \;\; q_{S_1}\ge 0 \;\;
\Leftrightarrow \;\; q_{S_2}\ge 0 \,.
\eqlabel{stability}
\end{equation}
$Q$ is stable if the only triangles for which $q_T=1$ are those 
with $Q_1$ or $Q_2$ equal to $Q$ (and the other equivalent to $0$).

Here $q_T$ is defined by the condition
\begin{equation}
ET + R_2T-TR_1 = q_T T
\end{equation}
where $R_1$ and $R_2$ are the R-matrices of $Q_1$ and $Q_2$,
respectively.

We can give one simple check that relates R-stability to a 
stability condition in the sense of Bridgeland \cite{bridgeland}.
Recall that in the orbifolded case (subsection \ref{orbifold}), 
we have defined morphism between objects in $\MF(W)$ by 
$\Hom^0(Q,Q') = \HH^{q=\varphi'-\varphi}(Q,Q')$. Therefore,
our condition \eqref{stability} directly implies
\begin{equation}
\varphi>\varphi' \Rightarrow \Hom^0(Q,Q')=0 \,,
\end{equation}
which is one of the axioms of \cite{bridgeland}. 

We also note that our formulation is similar to those of a stability 
condition for Lagrangian submanifolds proposed by Thomas \cite{thomas}
and further studied in \cite{thya}.

\subsection{A moment map problem?}
\label{moment}

The stability condition we have proposed is physically well-motivated.
It only deserves its name, however, if it can be related to the
moduli space problem for matrix factorizations. In other words,
one would like to show that stable matrix factorizations have
nicely behaved orbits under the group of gauge equivalences.
As we have mentioned before, this group is the group of similarity
transformations 
\begin{equation}
G \cong\GL^+(2N,\calr) \cong \GL(N,\calr)\times\GL(N,\calr)
\eqlabel{G}
\end{equation}
acting as in \eqref{similarity}. Thus we have an algebraic group
acting on a linear space with a constraint. This problem is
quite similar to the one studied by King \cite{king}. 

In \cite{king}, the general setup of geometric invariant theory (GIT)
\cite{mumford} is used to define moduli spaces for representations 
of finite-dimensional algebras, which can be equivalently described 
as the representations of quiver diagrams. Quivers arise naturally 
as world-volume theories for D-branes at singularities, and the theory
of quivers has played in important role in the development of 
$\Pi$-stability \cite{dfr,aspinwall}. In the quiver case, the gauge
group $\calg$ is the product of general linear groups acting on the 
vector spaces at each node of the quiver. King uses GIT to give a 
geometric description of the algebraic quotient of the representation 
space $Y$ with respect to a character $\chi:\calg\to\complex^\times$
via ``Mumford's numerical criterion'': A representation $y\in Y$ is
$\chi$-semistable iff $\chi$ is trivial on the stabilizer of $y$ and
if every one-parameter subgroup $g(\lambda)=\ee^{\lambda a}$
of $\calg$, for which $\lim_{\lambda\to\infty} y$ exists, satisfies
$\langle d\chi,a \rangle \ge 0$, where $d\chi$ is the infinitesimal
version of $\chi$ evaluated on the generator $a$ of $g(\lambda)$.

Our stability condition is precisely equivalent to such a ``numerical 
criterion''. To see this, note that all triangles in $\mf(W)$ are 
isomorphic to the standard cone \eqref{standard}, namely 
\begin{equation}
Q = Q_1\oplus Q_2 +  T \,,
\qquad 
R = R_1 \oplus R_2 + (q_t-1) \biggl[\frac{N_2}{N_1+N_2}S_1-
\frac{N_1}{N_1+N_2} S_2\biggr] \,,
\end{equation}
where $S_i=\id_i$. Under the one-parameter group of gauge 
transformations generated by $V=S_1$, this cone transforms as
\begin{equation}
Q_\lambda = \ee^{\lambda V} Q \ee^{-\lambda V} =
Q_1\oplus Q_2 + \ee^{-\lambda} T \underset{\lambda\to\infty}
{\longrightarrow}
Q_1\oplus Q_2 \,.
\eqlabel{split}
\end{equation}
The limit $\lambda\to\infty$ simply splits the cone back into its
constituents. The condition $q_T\le 1$ is equivalent to
\begin{equation}
- \tr (RV) = - (q_t-1)\frac{2 N_2 N_1}{N_2+N_1}  \ge 0 \,,
\end{equation}
thus identifying $\tr(R\,\cdot\,)$ as the character of $G$ with respect 
to which we are defining stability. The condition \eqref{condition} we 
are imposing to fix the ambiguities of $R$ is precisely the condition 
that $\tr(R\,\cdot\,)$ should vanish on the trivially acting gauge 
transformations. Similarly as in \cite{king}, we can then formulate 
a ``numerical criterion'' that a matrix factorization $Q$ is R-semistable 
if all one-parameter subgroups $\ee^{\lambda V}$ of the gauge group,
for which the limit $\lim_{\lambda\to \infty} \ee^{\lambda V}Q 
\ee^{-\lambda V}$ exists, satisfy $\tr(R V)\le 0$.

In \cite{king}, King then goes on to describing a symplectic quotient 
construction of the moduli space, which is the basis for the relation 
to quiver gauge theories.

We can at present see two difficulties in making such a relation in
our situation more precise, both of which due to the fact that 
$G$ is not as simple a gauge group as the one acting on quiver 
representations. First of all, as a complex Lie group, $G$ is 
infinite-dimensional. This is similar to the situation with vector 
bundles or Lagrangian submanifolds, giving reason for hope. The second 
difficulty appears if, as might seem natural, we would restrict to 
the degree $0$ gauge transformations, \ie, those generated by
\begin{equation}
\gg^0 = \bigl\{V\in\gg; EV+[R,V] = 0 \bigr\} \,.
\end{equation}
The problem with $\gg^0$ is that it is non-reductive. Indeed, since 
both polynomial and total degree are preserved in matrix multiplication, 
$\gg^0$ has a maximal solvable subalgebra consisting of those matrices 
without constant term. In other words, we can decompose $\gg^0$ into
its maximal reductive subalgebra consisting of those elements annihilated
by $E$, and the nilpotent part. This non-reductiveness of $\gg^0$
makes it more difficult to apply the general results of GIT and to 
find a relation with a moment map problem. In any case, however,
restricting to gauge transformations of degree $0$ makes the 
description of brane antibrane annihilation somewhat unnatural,
see subsection \ref{bbar}.

Setting aside these difficulties for the moment, we will naively 
follow the usual steps to write down a moment map-like flow 
equation on the gauge orbits. As we will see in section \ref{examples}, 
this naive flow works quite well in a number of examples. 
Imitating \cite{king}, we introduce a metric on the space of
matrix factorizations,
\begin{equation}
\langle Q,Q'\rangle = \sum_{\alpha} \tr \bigl(Q_\alpha^\dagger 
Q'_\alpha\bigr) \,,
\end{equation}
where 
\begin{equation}
Q= \sum_{\alpha} Q_\alpha x^\alpha \,,
\qquad\qquad
Q' = \sum_{\alpha} Q'_\alpha x^\alpha
\end{equation}
is the decomposition of $Q$ and $Q'$ into a sum over monomials 
$x^\alpha=\prod_i x_i^{\alpha_i}$. In any given case, we restrict to a 
finite-dimensional subgroup of $G$ and choose a basis of generators, 
$\{V_i\}$. The flow equation then is
\begin{equation}
\frac{dQ}{dt} = - \bigl(\langle Q, [V^i,Q]\rangle -
\tr R V^i\bigr) [V_i,Q] \,.
\eqlabel{flow}
\end{equation}
Note that by construction, \eqref{flow} is indeed a moment map for
the maximal reductive subgroup of the degree $0$ gauge group.

Moreover, one can see that the flow \eqref{flow} indeed reproduces the 
correct splitting \eqref{split} of the standard cone \eqref{standard} 
in the case that $q_T\ge 1$. A simple calculation gives
\begin{equation}
\frac{d\lambda}{dt} = \bigl(\ee^{-2\lambda} ||T||^2 + 
(q_T-1) \beta \bigr)
\eqlabel{ofcourse}
\end{equation}
where $\beta=2N_2N_2/(N_2+N_1)$. Evidently, this has a solution at finite
$\lambda$ if $q_T<1$, whereas for $q_T\ge 1$, the flow drives us to 
$\lambda\to\infty$. The form of eq.\ \eqref{ofcourse} is of course 
familiar in the context of solving D-flatness conditions in 
four-dimensional $\caln=1$ supersymmetric gauge theories.

\section{Examples}
\label{examples}

We conclude the paper with several concrete examples of matrix 
factorizations and flows on their gauge orbits defined by 
\eqref{flow}. As alluded to before, one can view these flows as toy 
models for boundary flows in Landau-Ginzburg models. (Landau-Ginzburg 
descriptions of boundary flows have also recently been discussed in 
\cite{cappelli}.)

\subsection{Minimal models}

Matrix factorizations of A-type minimal models with type 0A GSO
projection, corresponding to the LG superpotential $W=x^h$ were 
discussed detail in \cite{bhls,kali3,hor2,hll2}. They are given by
\begin{equation}
Q_n = \begin{pmatrix} 0 & x^n \\ x^{h-n}&0\end{pmatrix}
\qquad\text{for $n=1,\ldots,h-1$\,.}
\end{equation}
The R-matrix is 
\begin{equation}
R = \begin{pmatrix} \frac12-\frac nh& 0 \\
0 & -\frac12 + \frac nh
\end{pmatrix} \,.
\end{equation}
It is easy to see that there is only one non-trivial element
of the degree $0$ gauge algebra,
\begin{equation}
V = \begin{pmatrix} \frac12 & 0 \\ 0 & -\frac12 \end{pmatrix} \,.
\end{equation}
The orbit generated by $V$ looks like
\begin{equation}
Q_n(\lambda)= \ee^{\lambda V} Q_n \ee^{-\lambda V}
= \begin{pmatrix} 0 & \ee^{\lambda} x^n \\
\ee^{-\lambda} x^{h-n}\end{pmatrix} \,,
\end{equation}
so that the flow \eqref{flow} becomes
\begin{equation}
\begin{split}
\frac{d\lambda}{dt} & = -\bigl(\langle Q_n(\lambda),
[V,Q_n(\lambda)]\rangle - \tr R V\bigr) \\
&=  - \bigl(\ee^{2\Re\lambda} - \ee^{-2\Re\lambda}
- \frac12+\frac{n}h\bigr)
\end{split}
\eqlabel{mimoflow}
\end{equation}
Obviously, this flow has just one stationary point, which is stable.
This is as expected. Indeed, on can check explicitly that
all open strings between different minimal model factorizations 
satisfy $0\le q \le \hat c=1-\frac2h<1$.

\subsection{Brane-antibrane annihilation}
\label{bbar}

We have been tempted several times in this paper to restrict attention
to the gauge transformations of degree $0$ only. In this subsection, we 
show that in fact, the description of brane-antibrane annihilation in the 
context of matrix factorizations requires the inclusion of gauge 
transformation of non-zero degree.

Let $(f,g)$ be a matrix factorization of $W$ with
R-matrix $R = (R_+,R_-)$, and consider the cone over the 
identity $\id:(f,g)\to (f,g)$,
\begin{equation}
Q_0 = \begin{pmatrix} 
0 & 0 & f & 0 \\
0 & 0 & 1 & g \\
g & 0 & 0 & 0 \\
-1 & f & 0 & 0 
\end{pmatrix} \,.
\end{equation}
This is gauge equivalent to direct sums of the trivial 
factorization $W = 1 \cdot W$ via the gauge transformation
\begin{equation}
U_\lambda = \begin{pmatrix} 
1 & -\lambda f & 0 & 0 \\
0 & 1 & 0 & 0 \\
0 & 0 & 1 & \lambda g\\
0 & 0 & 0 & 1
\end{pmatrix} 
\end{equation}
Namely,
\begin{equation}
Q_\lambda = U_\lambda Q_0 U_\lambda^{-1} 
= \begin{pmatrix}
0 & 0 & (1-\lambda) f & (\lambda^2 -2\lambda)W\\
0 & 0 & 1 & (1-\lambda) g \\
(1-\lambda) g & (2\lambda-\lambda^2) W & 0 & 0 \\
-1 & (1-\lambda) f & 0 & 0
\end{pmatrix}
\eqlabel{bbarorb}
\end{equation}
which for $\lambda=1$ becomes
\begin{equation}
Q_1=
\begin{pmatrix}
0 & 0 & 0 & -W \\
0 & 0 & 1 & 0 \\
0 & W & 0 & 0 \\
-1 & 0 & 0 & 0
\end{pmatrix}
\end{equation}
What is the R-matrix associated with $Q_0$? The cone construction of 
subsection \ref{conegrad} gives one possible solution \eqref{choose}
\begin{equation}
R^{\rm cone} = \begin{pmatrix}
R_+ -\frac12 & 0 & 0 & 0 \\
0 & R_-+\frac12 & 0 & 0 \\
0 & 0 &  R_--\frac12 & 0 \\
0 & 0 & 0 & R_++\frac12
\end{pmatrix}
\end{equation}
However, this R-matrix does not satisfy \eqref{condition}.
By using the equivalence with $Q_1$, one finds that the 
generators of $C^0(Q_0)$ with non-vanishing diagonal entries 
and degree $0$ with respect to $R^{\rm cone}$ are
\begin{equation}
V_i = \begin{pmatrix}
e_i & - f_i & 0 & 0 \\
0 & 0 & 0 & 0 \\
0 & 0 & 0 & -g^i \\
0 & 0 & 0 & e_i
\end{pmatrix}
\quad
\text{and}
\qquad
V^i = \begin{pmatrix}
0 & f^i & 0 & 0 \\
0 & e_i & 0 & 0 \\
0 & 0 & e_i & g_i \\
0 & 0 & 0 & 0
\end{pmatrix}
\end{equation}
where $e_i$ is the $N\times N$ matrix with a $1$ at the $i$-th 
position on the diagonal, and zeroes elsewhere, and $f_i=e_i f$ 
and $f^i=f e_i$ are the $i$-th row and column of $f$, 
respectively. 

Then the combination of $R^{\rm cone}$, $V_i$ and $V^i$ 
satisfying \eqref{condition} is
\begin{equation}
R_0 = 
\begin{pmatrix}
-\frac12 & R_+f-fR_- & 0 & 0 \\
0 & \frac12 & 0 & 0 \\
0 & 0 & -\frac12 & g R_+-R_-g \\
0 & 0 & 0 & \frac12
\end{pmatrix}
=
\begin{pmatrix}
-\frac12 & -Ef +f & 0 & 0\\
0 & \frac12 & 0 & 0\\
0 & 0 & -\frac12 & Eg-g\\
0 & 0 & 0 & \frac12
\end{pmatrix}
\end{equation}
Under the similarity transformation \eqref{bbarorb}, $R_0$
transforms into the diagonal matrix $R_1={\rm diag}(-1/2,1/2,
-1/2,1/2)$. This $R_1$ is the R-matrix one would naturally assign 
to a sum of copies of the trivial branes described by $Q_1$.

Note that while the gauge transformation relating $Q_0$ and
$Q_1$ has degree zero with respect to $R^{\rm cone}$, it does not
have definite degree with respect to $R_0$. We conclude that
either are we forced to work with gauge transformations of
non-zero degree or we should be using $R^{\rm cone}$ as
R-matrix for the cone. We cannot completely exclude the 
second possibility since (by definition) the factorization 
$(1,W)$ does not have any non-trivial morphism ending on it,
so there are no R-charges to check. But the symmetric 
end-result, $R_1$, is good justification for the procedure
we have proposed. And the moment map equation only makes sense 
if we use $R_0$. One can also check that the flow defined 
by \eqref{flow} on the gauge orbit \eqref{bbarorb} flows to 
$\lambda=1$.

\subsection{Boundary flows in minimal models}

Having argued for the general relevance of gauge transformations of
non-zero degree, we now return to minimal models and study boundary 
flows associated with perturbations by a boundary condition changing 
operator. General aspects of such boundary flows in $\caln=2$ 
minimal models were discussed recently in \cite{hor2} and in 
\cite{hll2}. In particular, these works discuss the similarity 
transformations relating the different minimal model branes at 
the topological level, as well as the operators inducing these 
relations. Our flow equation \eqref{flow} gives a handle on the 
complete flow in the physical theory.

We will consider as an example the starting point
\begin{equation}
Q_0 = \begin{pmatrix} 0 &  0 & x^2 & 0 \\
0 & 0 & -x & x^3 \\
x^{h-2} & 0 & 0 & 0 \\
x^{h-4} & x^{h-3} & 0 & 0
\end{pmatrix}
\eqlabel{bfmm}
\end{equation}
The R-matrix, obtained by methods as above is
\begin{equation}
\begin{pmatrix}
\frac12-\frac 4h & -\frac xh& 0 & 0\\
0 & \frac 12 -\frac1h & 0 & 0\\
0 & 0 & -\frac12+\frac1h & -\frac{x^2}{h} \\
0 & 0 & 0 & -\frac12+\frac4h
\end{pmatrix}
\end{equation}
We have studied the flow induced by \eqref{flow} on the orbit
of $Q_0$ under the gauge transformations generated by
\begin{equation}
V= 
\begin{pmatrix}
\lambda_1 & \lambda_3 x & 0 & 0\\
0 & \lambda_2 & 0 & 0 \\
0 & 0 & -\lambda_2 & \lambda_4 x^2 \\
0 & 0 & 0 & -\lambda_1
\end{pmatrix}
\eqlabel{vmm}
\end{equation}
and find that it does converge to the diagonal
\begin{equation}
\begin{pmatrix}
0 & 0 & 0 & \alpha x^4 \\
0 & 0 & - \beta x & 0 \\
0 & - \beta^{-1} x^{h-1} & 0 & 0\\
\alpha^{-1} x^{h-4} & 0 & 0 & 0
\end{pmatrix}
\cong Q_1\oplus Q_4 \,,
\end{equation} 
where $\alpha$ and $\beta$ are the appropriate solutions of 
\eqref{mimoflow}. The R-matrix becomes ${\rm diag}(1/2-4/h,
1/2-1/h,-1/2+1/h,-1/2+4/h)$, which is certainly the correct 
result for this factorization.

We should note that the perturbation we have turned
on in \eqref{bfmm} is not the most relevant between the two minimal
model branes $Q_2$ and $Q_3$. We have chosen this one to illustrate
that there are various possible flow patterns in $\caln=2$ minimal
models. In this simple case the range of possibilities is essentially 
governed by the K-theory, isomorphic to $\zet_H$. One can create 
a free K-theory (and make the perturbation in \eqref{bfmm} the most 
relevant one) by considering an appropriate orbifold.

In any case, the end-result of the flow is consistent with the 
predictions made, for instance, in \cite{hor2,cappelli}.

\subsection{D0-brane in quintic Gepner model}
\label{d0brane}

In \cite{aad}, matrix factorizations were constructed which describe
(at the topological level) D0-branes at the Landau-Ginzburg orbifold 
point in the K\"ahler moduli space of the quintic Calabi-Yau. 
We here want to address the issue whether these factorizations 
can be stable by checking that the open strings stretched between 
this D0-brane and the rational tensor products of minimal model branes 
satisfy the unitarity bound. While this does of course not settle
the question whether the D0-brane can become unstable far away from
the large volume limit, it is certainly a non-trivial check.

The superpotential of interest is $W=\sum_{i=1}^5 x_i^5$. The 
factorizations of \cite{aad} are tensor products of minimal model 
factorizations in three of the five minimal factors together 
with a non-factorisable factorization in the remaining two factors.
Since taking tensor products simply adds $\U(1)$ charges, but does 
not affect the unitarity bound, it will suffice to consider this two-variable
factorization. We can factorize
\begin{equation}
x^5-y^5 = (x-y) (x^4+x^3 y+x^2y^2+xy^3+y^4) \,.
\eqlabel{diaco}
\end{equation}
One can see that the R-matrix associated with this factorization
is ${\rm diag}(3/10,-3/10)$. We want to compute the charges of open 
strings between this factorization and the tensor product of minimal 
model branes
\begin{equation}
f = \begin{pmatrix} x & -y \\ -y^4 & x^4 \end{pmatrix} \,,
\qquad 
g = \begin{pmatrix} x^4 & y \\ y^4 & x \end{pmatrix}\,,
\qquad
R={\rm diag}(3/5,-3/5,0,0)\,.
\eqlabel{tenpro}
\end{equation}
As computed in \cite{aad}, there is one bosonic and one fermionic cohomology 
class between \eqref{diaco} and \eqref{tenpro}, represented by
\begin{equation}
\begin{split}
\Phi^0 &= \begin{pmatrix}
y^3 & 1 & 0 & 0\\
0 & 0 & y^3 & x^3+x^2y+xy^2+y^3
\end{pmatrix} \\[.2cm]
\Phi^1 &= \begin{pmatrix}
0 & 0 & -1 & 1\\
x^3+x^2 y+xy^2+y^3 & -1 & 0 & 0
\end{pmatrix} \,,
\end{split}
\end{equation}
respectively. We easily find
\begin{equation}
q(\Phi^0) = \frac{9}{10} \,, \qquad 
q(\Phi^1) = \frac{3}{10}
\end{equation}
satisfying the unitarity bound $0\le q\le \hat c=\frac65$.
We have also checked the open strings between the D0-brane
factorization and the other minimal model branes. They all
satisfy the bound.

\subsection{Decay of an unstable factorization}
\label{unsex}

In this subsection, we give an example of a matrix factorization
that is unstable and investigate to what extent our flow \eqref{flow}
can detect this without having to check the charges of open strings.
The superpotential is $W=x^5+y^5$, and the factorization given
by
\begin{equation}
\begin{split}
f_{\rm unst}=\begin{pmatrix} x & y & 0 \\
0 & x^3 & y \\
y^3 & 0 & x
\end{pmatrix} \qquad
g_{\rm unst} = 
\begin{pmatrix} 
x^4 & -x y & y^2 \\
y^4 & x^2 & - x y\\
-x^3 y^3 & y^4 & x^4 
\end{pmatrix}
\\
R={\rm diag}(7/10, -1/10, -1/10, 1/10, 1/10, -7/10)
\quad\;\;
\end{split}
\eqlabel{tbt}
\end{equation}
A morphism between this factorization and the tensor product of minimal
model branes \eqref{tenpro} is given by
\begin{equation}
T = \begin{pmatrix}
1& 0& 0& 0& 0& 0\\ 
0& x& -y& 0& 0& 0 \\
0& 0& 0& 1& 0& 0 \\ 
0& 0& 0& 0& 1& 0
\end{pmatrix}
\end{equation}
and one can easily check that this field has R-charge $-\frac1{10}$, 
violating the unitarity bound. (Note that since $T$ has scalar entries, 
it cannot be exact.) The cone over $T$ is another copy of the tensor 
product of minimal branes, thus exhibiting \eqref{tbt} as an unstable 
bound state, obtained by ``condensing'' a field with $q>1$ between two 
such objects. Namely, $(f_{\rm unst},g_{\rm unst})$ is stably equivalent
to
\begin{equation}
F_{\rm unst} =
\begin{pmatrix}
x & y & 0 & 0 \\
-y^4 & x^4 & 0 & 0 \\
0 & -x^3 & x^4 & -y \\
y^3 & 0 & y^4 & x 
\end{pmatrix}\,, \qquad
\text{with corresponding $G_{\rm unst}$\,.}
\end{equation}
To be precise, we should note that $(F_{\rm unst},G_{\rm unst})$ is
really a cone over a brane and its own antibrane, but via a field
that is not the identity. As a consequence, $(f_{\rm unst},
g_{\rm unst})$ has a unitarity violating field in the spectrum 
with itself. But since this field can easily be projected out
by going to an appropriate orbifold, it should not be viewed as
the cause of the instability.

In discussing the flow, it is useful to contrast the unstable factorization
with a very similarly structured stable (with the same caveat as before)
bound state of two minimal model tensor products, namely
\begin{equation}
F_{\rm stab} =
\begin{pmatrix}
x & y^2 & 0 & 0 \\
-y^3 & x^4 & 0 & 0 \\
0 & -x^3 & x^4 & -y^2 \\
y & 0 & y^3 & x 
\end{pmatrix}
\,, \qquad G_{\rm stab}\,,
\end{equation}
which can be reduced to
\begin{equation}
f_{\rm stab} = \begin{pmatrix}
x & y^2 & 0 \\
0 & x^3 & y^2\\
y & 0 & x 
\end{pmatrix} \,,
\qquad
g_{\rm stab} = {\rm adj}(f_{\rm stab}) \,.
\end{equation}

We have studied numerically the flow defined by \eqref{flow} on the
12-parameter gauge orbit of $(F_{\rm unst},F_{\rm unst})$ and 
$(F_{\rm stab},G_{\rm stab})$ generated by
\begin{equation}
\begin{pmatrix}
\lambda_1 & 0 & 0 & 0 & 0 & 0 & 0 & 0 \\
0 & \lambda_2 & x \lambda_3 & y\lambda_4 & 0 & 0 & 0 & 0  \\
0 & 0 & \lambda_5 & 0 & 0 & 0 & 0 & 0  \\
0 & 0 & 0 & \lambda_6 & 0 & 0 & 0 & 0  \\
0 & 0 & 0 & 0 & \lambda_7 & 0 & y\lambda_8 & 0  \\
0 & 0 & 0 & 0 & 0 & \lambda_9 & x\lambda_{10} & 0  \\
0 & 0 & 0 & 0 & 0 & 0 & \lambda_{11} & 0  \\
0 & 0 & 0 & 0 & 0 & 0 & 0 & \lambda_{12}  \\
\end{pmatrix}
\end{equation}
and
\begin{equation}
\begin{pmatrix}
\lambda_1 & 0 & 0 & 0 & 0 & 0 & 0 & 0 \\
0 & \lambda_2 & x \lambda_3 & y^2\lambda_4 & 0 & 0 & 0 & 0  \\
0 & 0 & \lambda_5 & 0 & 0 & 0 & 0 & 0  \\
0 & 0 & 0 & \lambda_6 & 0 & 0 & 0 & 0  \\
0 & 0 & 0 & 0 & \lambda_7 & 0 & y^2\lambda_8 & 0  \\
0 & 0 & 0 & 0 & 0 & \lambda_9 & x\lambda_{10} & 0  \\
0 & 0 & 0 & 0 & 0 & 0 & \lambda_{11} & 0  \\
0 & 0 & 0 & 0 & 0 & 0 & 0 & \lambda_{12}  \\
\end{pmatrix}
\end{equation}
respectively. We find that starting from quite general initial 
conditions, \eqref{flow} indeed drives $(F_{\rm unst},G_{\rm unst})$
to the split into the direct sum of two copies of the tensor
product brane. On the other hand $(F_{\rm stab},G_{\rm stab})$
flows to the direct sum of $(f_{\rm stab},g_{\rm stab})$ and
a copy of the trivial factorization $(1,W)$.

It is worthwhile emphasizing that this statement does not hold for 
all initial conditions. One of the consequence of non-reductiveness
is that the flow defined by \eqref{flow} is not convex. Taking
a second derivative on the right hand side does not produce
something positive definite because $\langle Q,[V,Q']\rangle
\neq \langle [V^\dagger,Q],Q'\rangle$ in general. If the flow
is not convex, there is no guarantee that stationary points will
be unique. In the present case, there does exist a stationary 
point for the flow close to $(f_{\rm unst},g_{\rm unst}\oplus 
(1,W)$. But this point is a saddle point of the flow, \ie, it 
is unstable in the sense of dynamical systems. One has to
account for this possibility if one wants to make sense of
\eqref{flow} in general.

Another example of the same character as the one we have been 
discussing in this subsection arises in the series of models 
$W=x^h+y^3$, with factorizations given by
\begin{equation}
f = \begin{pmatrix}
x^n & y & 0 \\
0 & x^n & y \\
y & 0 & x^{h-2n} 
\end{pmatrix}
\eqlabel{last}
\end{equation}
(and, as by now familiar, $g={\rm adj}(f)$). By considerations
similar to those we have given above, one finds that the factorizations
\eqref{last} are stable when $n< h/6$ and (apparently) stable otherwise.
(In particular, for $h< 6$, where $W$ describes an $E$-type minimal
model, these factorizations are all stable.)

\subsection{A non-homogeneous factorization}
\label{nonhomo}

Lest we leave the impression that all matrix factorizations
of quasi-homogeneous polynomials are quasi-homogeneous,
here is a counter-example. 

The superpotential $W=x^3+y^7$ is one of the simplest
superpotentials that is not a simple singularity. In fact,
it is unimodular. Torsion free rank one modules over 
local rings of unimodular singularities were classified in 
\cite{schappert}. It is a simple exercise to determine the 
associated matrix factorizations. On the list for $W=x^3+y^7$, 
one finds the following one-parameter family of factorizations.
\begin{equation}
\begin{split}
f &= \begin{pmatrix} x^2-\lambda y^5 & x y\\ xy+\lambda^2 y^4
& -\lambda x+y^2
\end{pmatrix} \\
g &= \begin{pmatrix} x -\frac{y^2}\lambda &
\frac{x y }{\lambda} \\ \frac{xy}{\lambda}+
\lambda y^4 & -\frac{x^2}{\lambda}+y^5
\end{pmatrix} \,.
\end{split} 
\eqlabel{family}
\end{equation}
For $\lambda\neq 0$, this is stably equivalent to the following
factorization
\begin{equation}
\begin{split}
\tilde f &=
\begin{pmatrix}
-x y^5 & \lambda x y^4 + y^6 & -x^2 \\
y^6 & x^2-\lambda y^5 & x y\\ 
-x^2-\lambda y^5 & y x + \lambda^2 y^4 & -\lambda x+y^2
\end{pmatrix}\\
\tilde g &= 
\begin{pmatrix} \lambda & y & -x \\ 
y & x & 0 \\ -x & \lambda y^4 & y^5
\end{pmatrix} 
\end{split} \,,
\eqlabel{tfamily}
\end{equation}
which is non-reduced, but has a limit as $\lambda\to 0$. While $W$ is 
quasi-homogeneous with $q_x=2/3$, $q_y=2/7$, we see that
\begin{equation}
E f -f + R_+ f - f R_- 
=
\frac{2}{21}\lambda
\begin{pmatrix}
-y^5 & 0 \\ 2 \lambda y^4 & -x 
\end{pmatrix} 
= \frac 2{21}\lambda \del_\lambda f \,,
\end{equation}
where 
\begin{equation}
R = \begin{pmatrix} R_+ & 0 \\0 & R_-\end{pmatrix}
=\begin{pmatrix} -\frac 7{42} & 0 & 0 &0\\0&\frac9{42}&0&0\\
0&0&\frac 7{42}&0\\0&0&0&-\frac 9{42}\end{pmatrix} \,.
\end{equation}
Since this $\del_\lambda f$ is the marginal deformation of
the family \eqref{family}, it is a non-trivial cohomology class
(this can also be checked directly). As a consequence,
the matrix factorization $(f,g)$ is not quasi-homogeneous.

It is interesting to ask for a geometric interpretation 
of this example. For example, one could embed $W=x^3+y^7$ into 
the appropriate Calabi-Yau Landau-Ginzburg model, and try to 
identify a mirror Lagrangian cycle. Non-homogeneity of $(f,g)$ 
should be mirror to non-vanishing of the Maslov class. 
Of course, it is not clear to what extent Lagrangians with 
non-vanishing Maslov anomaly participate in mirror symmetry. 
The intriguing point is that the limit of \eqref{family}
for $\lambda\to 0$ is actually quasi-homogeneous and therefore
might have a good mirror. One way to avoid the paradox 
conclusion that the deformation of a non-anomalous Lagrangian 
is anomalous would be to show that the brane described by
this factorization is never stable on the moduli space. (It
is unstable in the Landau-Ginzburg model, as the examples
in subsection \ref{unsex}.)

\section{Summary}
\label{summary}

For convenience and definiteness, we shall here give a summary
of the main ingredients that are proposed to enter into a
stability condition for matrix factorizations.

As explained in section \ref{review}, the physical origin of 
stability conditions in string theory is the grading by 
worldsheet R-charge. In the context of matrix factorizations, 
which originate in local commutative algebra, it is also quite 
natural to consider the graded situation, so it would not seem
that physics has much input to give. Before repeating the
claim that it does, it is worthwhile to fix the convenient
normalization of the grading: Physics suggests a normalization 
in which $W$ has charge $2$, giving the field variables fractional 
charge, whereas a more standard mathematical choice is to make all 
degrees integer. 

With this in mind, we have associated to any matrix factorization
$$
Q^2 = W
$$
of a Landau-Ginzburg potential $W$, satisfying the anomaly-free
condition that $EQ-Q$ is cohomologically trivial, a matrix $R$, 
defined by the conditions \eqref{cohomological} and \eqref{condition},
$$
E Q -Q = [Q,R] \qquad {\rm and}\qquad \tr(R V) = 0 \;\;\text{whenever 
$[V,Q]=0$,}
$$
where we have argued that the latter condition would fix $R$ uniquely.
Via \eqref{morgrad}, this induces a grading, $q$, of the morphism spaces.

The choice of normalization of the grading is important because we
intend to compare the grading of morphisms with another natural
quantity that can be associated to a Landau-Ginzburg potential,
namely the central charge
$$
\hat c \,.
$$
A mathematical quantity that is closely related to $\hat c$ is the 
so-called ``singular index'' that appears in singularity theory (see 
 \cite{arnold}), but it does not seem to have played a crucial 
role in the purely algebraic context so far.

The basic idea, motivated by $\Pi$-stability as we have explained,
is to impose the {\it unitarity constraint} \eqref{unitarity}
$$
0\le q \le \hat c
$$
as a stability condition on the category of topological D-branes.

The problem at this point, which is inherited from $\Pi$-stability, 
is that it is not {\it a priori} clear exactly {\it how} to impose this
condition. For example, should it be imposed on {\it all} morphisms, or only 
on all morphisms involving stable objects? Or should one rather attempt 
to define the stable branes as a ``maximal set'' of objects satisfying 
this (and maybe some other) condition? Although the latter option would 
seem to depend on too many arbitrary choices, such ambiguities might 
not be unnecessary. The set of stable objects is expected to be unique 
only up to auto-equivalences of the topological category or monodromies 
in the moduli space \cite{douglas,aspreview}.

In the mathematical approach of \cite{bridgeland}, the problem is
circumvented by postulating the existence of abelian subcategories
at each point in moduli space, on which a stability condition
can be imposed in a more standard well-defined form.

We have argued here that there should be a way to identify uniquely 
a set of stable objects at the Landau-Ginzburg point, essentially 
because our definition of the grading {\it does not depend} on the 
rest of moduli space, and is hence insensitive to monodromies. A 
posteriori, this should also provide an abelian subcategory.

To gain further confidence that such an approach is possible, we
have then proposed a relation to a moduli space problem via
a ``moment map-like'' flow equation \eqref{flow}
$$
\frac{dQ}{dt} = - \bigl(\langle Q, [V^i,Q]\rangle -
\tr R V^i\bigr) [V_i,Q] \,,
$$
which is expected to provide the split of any given object into
its stable constituents. We have implemented this flow in various 
relevant examples, with reasonable results.

\begin{acknowledgments}
It is a pleasure to thank Mike Douglas, Kentaro Hori, Calin Lazaroiu, 
Dmitri Orlov, Gerhard Pfister, and Piljin Yi for useful discussions, 
comments, and correspondence. Parts of this work were originally presented 
at the Fields Institute Workshop on Mirror Symmetry held at Perimeter 
Institute, November 19-23, 2004. I would like to thank the organizers for 
the invitation to speak and the participants, especially Andrei
Caldararu, Charles Doran, and Anton Kapustin, for valuable feedback.
This work was supported in part by the DOE under grant number
DE-FG02-90ER40542.
\end{acknowledgments}

\end{document}